\documentclass[10pt,journal,compsoc]{IEEEtran}

\makeatletter
\def\ps@headings{%
\def\@oddhead{\mbox{}\scriptsize\rightmark \hfil \thepage}%
\def\@evenhead{\scriptsize\thepage \hfil \leftmark\mbox{}}%
\def\@oddfoot{}%
\def\@evenfoot{}}
\makeatother
\usepackage{cite}
\usepackage{color}
\usepackage{graphicx}
\usepackage{amssymb,wasysym}
\usepackage[vlined,linesnumbered,ruled,norelsize]{algorithm2e}
\usepackage[usenames,dvipsnames]{xcolor}

\usepackage{array}
\usepackage{subfigure}
\usepackage{float}
\usepackage[switch]{lineno}
\usepackage{xcolor}
\usepackage{url}

\renewcommand{\IEEEQED}{\IEEEQEDopen}
\newtheorem{theorem}{\textbf{Theorem}}

\newtheorem{lemma}{\textbf{Lemma}}

\begin{document}

\title{Truthful Scheduling Mechanisms for Powering Mobile Crowdsensing}

%\author{\IEEEauthorblockN{Kai Han\IEEEauthorrefmark{1}\IEEEauthorrefmark{2}~~~~~~~~Chi Zhang\IEEEauthorrefmark{1}~~~~~~~~Jun Luo\IEEEauthorrefmark{1}}\\
%\IEEEauthorblockA{\IEEEauthorrefmark{1}School of Computer Engineering, Nanyang Technological University, Singapore\\
%\IEEEauthorrefmark{2}School of Computer Science, Zhongyuan University of Technology, China\\
%%\IEEEauthorrefmark{2}Department of Computer Science and Technology, Tsinghua University, China\\
%Email: dr.kai.han@ieee.org, junluo@ntu.edu.sg, czhang8@e.ntu.edu.sg}}
\author{Kai~Han,~\IEEEmembership{Member,~IEEE,}
        Chi~Zhang~
        and~Jun~Luo,~\IEEEmembership{Member,~IEEE}% <-this % stops a space
\IEEEcompsocitemizethanks{\IEEEcompsocthanksitem Kai Han is with School of Computer Engineering, Nanyang Technological University, Singapore, and also with School of Computer Science, Zhongyuan University of Technology, China.\protect\\
%E-mail: dr.kai.han@ieee.org
\IEEEcompsocthanksitem Chi Zhang and Jun Luo are with School of Computer Engineering, Nanyang Technological University, Singapore.\protect\\
%E-mail: \{czhang8,junluo\}@ntu.edu.sg
}% <-this % stops a space
\thanks{}}

%\author{\IEEEauthorblockN{Michael Shell}
%\IEEEauthorblockA{School of Electrical and\\Computer Engineering\\
%Georgia Institute of Technology\\
%Atlanta, Georgia 30332--0250\\
%Email: http://www.michaelshell.org/contact.html}
%\and
%\IEEEauthorblockN{Homer Simpson}
%\IEEEauthorblockA{Twentieth Century Fox\\
%\and
%\IEEEauthorblockN{James Kirk\\ and Montgomery Scott}
%\IEEEauthorblockA{Starfleet Academy\\
%San Francisco, California 96678-2391\\
%\IEEEspecialpapernotice{A Submission to IEEE ICNP 2013}
\maketitle

\begin{abstract}
%\boldmath
  Mobile crowdsensing leverages mobile devices (e.g., smart phones) and human mobility for pervasive information exploration and collection; it has been deemed as a promising paradigm that will revolutionize various research and application domains. Unfortunately, the practicality of mobile crowdsensing can be crippled due to the lack of incentive mechanisms that stimulate human participation. In this paper, we study incentive mechanisms for a novel Mobile Crowdsensing Scheduling (MCS) problem, where a mobile crowdsensing application \textit{owner} announces a set of sensing tasks, then human \textit{users} (carrying mobile devices) compete for the tasks based on their respective sensing costs and available time periods, and finally the owner schedules as well as pays the users to maximize its own sensing revenue under a certain budget. We prove that the MCS problem is NP-hard and propose polynomial-time approximation mechanisms for it. We also show that our approximation mechanisms (including both offline and online versions) achieve desirable game-theoretic properties, namely truthfulness and individual rationality, as well as $\mathcal{O}(1)$ performance ratios. Finally, we conduct extensive simulations to demonstrate the correctness and effectiveness of our approach.
\end{abstract}

\IEEEpeerreviewmaketitle

\section{Introduction} \label{sec:intro}
  With the proliferation of palm-size mobile devices (smart phones, PDAs, etc.), we have a new tool for pervasive information collection, sharing, and exploration. For those information that traditionally require specific (possible very expensive) instruments or devices to gather, they can now be outsourced to human crowds.  Moreover, as this tool relies on human mobility and activity to bring their mobile devices around, it also introduces a new type of social action: mobile crowdsensing. Recently, there have emerged numerous systems based on this idea across a wide variety of research and application domains, such as healthcare, social networks, safety, environmental monitoring, and transportation~\cite{GantiYL2011,Khan2013}.

  Whereas mobile crowdsensing appears to be a promising paradigm that will revolutionize many research and application domains and ultimately impact on our everyday life significantly, it cannot take place spontaneously in practice, as many other social actions. Apparently, participating in a mobile crowdsensing task usually requires a mobile device carrier (\textit{users} hereafter) to move to specific areas where data gathering is required, to turn on his/her sensors for gathering data (e.g., GPS locations), and to upload the sensing data to an online server. These actions inevitably incur sensing costs in terms of, for example, (device) energy consumption/depreciation and Internet access. Therefore, from a pragmatic point of view, human crowds may not be willing to participate in mobile crowdsensing unless they are incentivized. Therefore, proper incentive mechanisms are crucial for enabling mobile crowdsensing, and this intriguing problem has started to attract attentions very recently~\cite{Lee2010,Jaimes2012,Duan2012,Yang2012,Qinghua2013}.

  Designing incentive mechanisms for mobile crowdsensing is challenging, particularly because the designer face rational but selfish users who can act \textbf{strategically} (i.e., lying about their private information) to maximize their own utilities. To handle this issue, a mechanism needs to motivate the users to report their real private information, or in game theoretical term, the mechanism should be \textit{truthful}~\cite{Nisan2007}. Certain existing proposals~\cite{Lee2010,Jaimes2012,Qinghua2013} do not take truthfulness into account for the designed incentive mechanisms. Such mechanisms, though being able to motivate user participation in a mobile crowdsensing application, may end up costing the application \textit{owner} a big fortune to obtain a certain sensing revenue. % jeopardizing the revenue of whose budge is limited.
  %and align their interests with the system goals; without compromising the optimality of the solution too much, while at the same time achieve computational efficiency.

  Mobile crowdsensing also imposes a \textbf{unique} requirement on incentive mechanism design, compared with conventional crowdsourcing applications (e.g., Amazon Mechanical Turk\footnote{\url{https://www.mturk.com/mturk/welcome}}). In particular, as most mobile crowdsensing tasks entail a certain level of temporal coverage of the sensing area \cite{GantiYL2011} whereas individual users have limited participating time, the private information pertaining to individual users includes not only the sensing costs but also the available time periods. This unique requirement makes mechanism design even more challenging since i) it faces multi-parameter environments where users' private information is multi-dimensional, ii) it has to schedule users properly for revenue maximization, and iii) it should be able to handle the dynamic arrivals of the users. To the best of our knowledge, these issues have never been tackled in the literature by far.

  In this paper, we investigate a novel scheduling problem arising from the mobile crowdsensing context, where an owner announces a set of sensing tasks with various values, and users with different available time and sensing costs bid for these tasks. We design mechanisms to schedule the users for maximizing the total sensing value obtained by the owner under a certain budget, while achieving multiple performance objectives including truthfulness, individual rationality~\cite{Nisan2007}, provable approximation ratios, and computational efficiency simultaneously. Moreover, our mechanisms work for both offline (users all arrive together) and online (user may arrive sequentially) cases. In summary, we have the following major contributions in our paper:
  \begin{itemize}
    \item We formally formulate the Mobile Crowdsensing Scheduling (MCS) problem and prove that it is NP-hard.
    \item We propose an offline polynomial-time mechanism for the MCS problem with $\mathcal{O}(1)$ approximation ratio, which is truthful given that the users strategically report their multi-dimensional private information including the sensing costs and available time periods.
    \item We propose an online polynomial-time mechanism for the MCS problem with $\mathcal{O}(1)$ competitive ratio, which is truthful if the users strategically report their sensing costs.
    \item We conduct extensive simulations, and the simulation results demonstrate the effectiveness of our approach.
  \end{itemize}
  %To the best of our knowledge, we are the first to study the mobile sensing scheduling problem and provides incentive mechanisms with guaranteed performance for it.

  The remaining of our paper is organized as follows. We introduce the models and assumptions in Sec.~\ref{sec:model}, where we also formulate the MCS problem. Then we first present an approximation algorithm for the MCS problem in Sec.~\ref{sec:appro}. Based on this algorithm, we further propose truthful mechanisms for the MCS problem under both offline and online settings in Sec.~\ref{sec:offline} and Sec.~\ref{sec:online}, respectively. We report the results of our extensive simulations in Sec.~\ref{sec:simu}. We finally discuss the related work in Sec.~\ref{sec:rw}, as well as conclude our paper in Sec.~\ref{sec:con}. In order to maintain fluency, we only prove a few crucial theorems in the main texts but postpone most of the (sketched) proofs to the Appendix.

\section{Modeling and Problem Formulation} \label{sec:model}
  We formally introduce the assumptions and definitions for the MCS problem in this section.

  \subsection{The Mobile Crowdsensing Scheduling Problem}
    We assume that a mobile crowdsensing application \textit{owner} announces a set of \textit{sensing tasks} $\mathcal{K}=\{K_1,K_2,...K_m\}$, and that performing any task $K_i: 1\leq i\leq m$ per unit time has a \textit{sensing value} $u_i\in \mathbb{R}^+$ to the owner, whereas performing any task for less than one unit time has a sensing value 0.
    %Following~\cite{Yang2012}, we assume that duplicate sensing does not provide extra value, i.e., if a sensing task is performed multiple times, the sensing value is same to that of performing the sensing task once.
    We also assume that the owner holds a \textit{budget} $G\in \mathbb{R}^+$: the maximum amount of total payment that it is willing to make for outsourcing the sensing tasks in $\mathcal{K}$ to others.

    Suppose that a set of \textit{users} (or sensor carriers), denoted by $\mathcal{A}=\{A_1,A_2,...,A_n\}$, may potentially perform the sensing tasks in $\mathcal{K}$. Each user $A_i$ is able to perform one sensing task $\kappa_i\in \mathcal{K}$, and has a \textit{private value} $\hat{d}_i\in \mathbb{R}^+$ indicating his/her sensing cost per unit time. For convenience, let $\mu_i=u_{\kappa_i}$ denote the value of $A_i$ to the owner for $A_i$'s one unit time sensing on task $\kappa_i$. We also assume that a user $A_i$ is only available during the time period $[\hat{s}_i,\hat{e}_i]$, where $\hat{s}_i,\hat{e}_i \in \mathbb{Z}$ are the earliest and latest available points in time private to $A_i$. Here $\hat{s}_i$ and $\hat{e}_i$ are both integers as they are defined with respect to certain time units. In reality, a user cannot be available all the time for sensing due to, for example, his/her own career. Therefore, we use an integer constant $\lambda$ to denote the upper bound of $(\hat{e}_i - \hat{s}_i), \forall i =  1,\cdots,n$. %the number of time units in any user's available time period.

\begin{figure}[htb]
      \centering
      \includegraphics[width=.88\columnwidth]{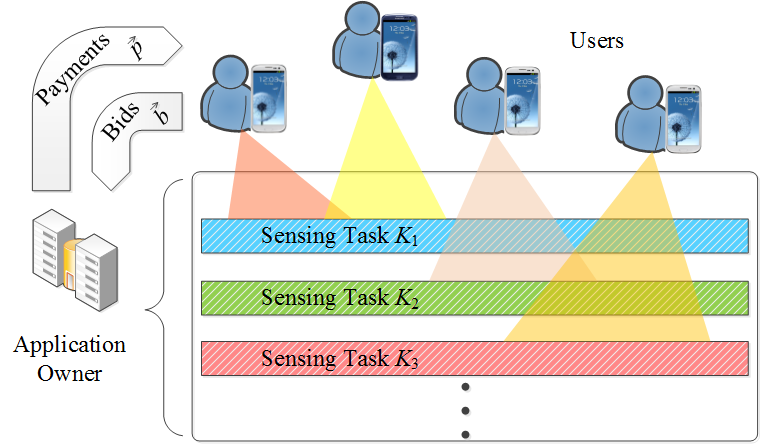}
      \caption{The mobile crowdsensing scheduling (MCS) problem.} \label{fig:mcsmodel}
      %\vspace{-1ex}
    \end{figure}

    In a Mobile Crowdsensing Scheduling (MCS) problem (briefly illustrated in Fig.~\ref{fig:mcsmodel}), the owner solicits the \textit{bids} $\vec{b}=(b_1,b_2,...,b_n)$ from the users in $\mathcal{A}$; each $b_i\!:\! 1\leq i\leq n$ is a 3-tuple $(d_i,s_i,e_i)$ where $d_i\in \mathbb{R}^+$ and $[s_i,e_i]$ ($s_i,e_i \in \mathbb{Z}$) are $A_i$'s declared sensing cost (per unit time) and available time period, respectively. Let $T_i = [s_i,e_i]$ for brevity. The owner then finds a \textit{sensing time schedule} $\vec{y}(\vec{b})=(y_1(\vec{b}),y_2(\vec{b}),...,y_n(\vec{b}))$ for the users, where $y_i(\vec{b})\subseteq T_i$ is the time period allocated to $A_i$ for sensing and it is not necessarily continuous. We also denote by $|y_i(\vec{b})|$ the \textit{total length} of $y_i(\vec{b})$ in time. Based on the bids, the owner also computes a \textit{payment vector} $\vec{p}(\vec{b})=(p_1(\vec{b}),p_2(\vec{b}),...,p_n(\vec{b}))$, where $p_i(\vec{b}) \ge 0$ is the payment to $A_i$ and $\sum_{i=1}^n p_i(\vec{b})\leq G$ should be satisfied. Moreover, the payment to any user $A_i$ should be no less than his/her total sensing cost if all users bid truthfully, i.e., $p_i(\vec{b})\geq d_i|y_i(\vec{b})|$. Defining the owner's \textit{revenue}, $R\left(\vec{y}(\vec{b})\right)$, as the total sensing value of performing the sensing tasks allocated by the sensing schedule $\vec{y}(\vec{b})$, i.e.,
    %to ensure \textit{voluntary participation} (a.k.a. \textit{individual rationality})~\cite{Fudenberg1991},
    %
    \begin{eqnarray}
      R\left(\vec{y}(\vec{b})\right) &=& \sum\nolimits_{i=1}^m u_i\cdot \left\lfloor\left|\bigcup\nolimits_{j:\kappa_j=i}y_j(\vec{b})\right|\right\rfloor, \nonumber
    \end{eqnarray}
    the goal of the MCS problem is to maximize this revenue subject to all the above constraints. Note that should we assume that the users always bid truthfully, the MCS problem would become a pure combinatorial optimization problem. As proved by \textit{\textbf{Theorem}~\ref{thm:nphard}}, this simplified problem is NP-hard.
    \begin{theorem}
      The MCS problem with all users bidding truthfully is NP-hard.
      \label{thm:nphard}
    \end{theorem}

    For notational simplicity, we sometimes omit $\vec{b}$ when writing the schedule and payment vectors (e.g., writing $\vec{y}$ instead of $\vec{y}(\vec{b})$), if the bid $\vec{b}$ is clear from the context.

  \subsection{Offline and Online Truthful Mechanisms}
    Each user $A_i$ has a \textit{utility} indicating the difference between the payment made to him/her and his/her total sensing cost according to the sensing schedule, i.e., $p_i(\vec{b})-\hat{d}_i|y_i(\vec{b})|$. In practice, the users are selfish and are only interested in maximizing their own utilities. For this purpose, they may bid strategically, i.e., lying about their private information such as sensing costs and available time. To handle users' strategic behaviors, we need to design \textit{truthful} mechanisms for the MCS problem to align the users' interests with the system goal of revenue maximization. A mechanism is called (dominant-strategy) truthful if any user maximizes his/her utility by revealing his/her real private information, no matter how other users may act~\cite{Nisan2007}. A randomized mechanism is called truthful (or \textit{universally-truthful}) if it is a randomization over a set of truthful mechanisms. Moreover, we also require our mechanisms to satisfy \textit{individual rationality} (IR here after), which means that any truth-telling user $A_i$ always gets a non-negative utility~\cite{Nisan2007}, i.e., $p_i((\hat{d}_i,\hat{s}_i,\hat{e}_i),b_{-i})\geq \hat{d}_i|y_i((\hat{d}_i,\hat{s}_i,\hat{e}_i),b_{-i})|$, where $b_{-i}$ represents the bids of the users other than $A_i$.

    %A mechanism is called \textit{truthful} if each user maximizes his utility by declaring his true sensing cost per unit time and true available sensing period irrespective of the bids of the other participants (i.e., truthful bidding is a \textit{dominant strategy} for each user, by the game theory jargon), and a randomized mechanism is called truthful (or \textit{universally-truthful}) if it is a randomization over truthful mechanisms.

    In the following sections, we aim to design truthful mechanisms for the MCS problem under both the offline and online settings. In the offline setting, the owner collects all the users' bids before scheduling them, which corresponds to a practical scenario that the users reserve sensing tasks in advance. In the online setting, the users' bids are revealed one by one, and the owner must make an irrevocable decision on scheduling any user right at the moment when the user's bid is revealed. This setting corresponds to another practical scenario where the users arrive randomly at some sensing area, and we assume that the users' arrival order is drawn uniformly at random from the set of all permutations over the users.

    %the authority have no knowledge about the users that are not arrived yet. In our case, this means that bidding $b_i=c_i$ is the best strategy for any $A_i$, and no participants can increase his utility by unilaterally changing his submitted bid. However, guaranteing truthfulness may compromise optimality.

\section{Approximation Algorithms for MCS} \label{sec:appro}
  In this section, we treat the MCS problem as a pure combinatorial optimization problem and design approximation algorithms for it, as shown in \textbf{Algorithm~\ref{alg:assc}}. Although the strategic behaviors of the users are not considered in \textbf{Algorithm~\ref{alg:assc}}, this algorithm serves as an important building block for the truthful mechanisms designed later.
  \begin{algorithm}[htb]
%   \KwIn{$G, \vec{b}, \mathcal{A}, \mathcal{T}, \beta$}
%   %
%   \KwOut{$\vec{w}(\vec{b})$}
    %
    $\mathcal{D}\leftarrow \mathcal{A}$;~~~$\mathcal{W} \leftarrow \emptyset$;~~~
    \lFor{$i\leftarrow 1$ \KwTo $n$}{
        $y_i(\vec{b})\leftarrow \emptyset$~\\
    }
    \Repeat {$q< |Z_j|$ or $\mathcal{D}= \emptyset$} {\label{ln:lpbegin}
        Find $j$ such that $A_j=\max_{\prec}(\mathcal{D})$ \label{ln:greedysel}~\\
%       $j\leftarrow \arg \max_{i:A_i\in D} \frac{\mu_i}{d_i}$ \\ \label{ln:greedysel}
        $Z_j\leftarrow T_j-\bigcup_{i:\kappa_i=\kappa_j}y_i(\vec{b})$\\ \label{ln:optionalperiod}
        $q\leftarrow \min\left\{|Z_j|, \left\lfloor\frac{G}{2d_j}-\frac{R(\vec{y}(\vec{b}))}{\mu_j}\right\rfloor\right\}$ \label{ln:allocatetime}~\\
        %\eIf{$\beta=1$}{
%            $q\leftarrow \min\{|Z_j|,  \frac{G-\sum_{i=1}^n d_i*|w_i(\vec{b})|}{d_j}\}$\\ \label{ln:allocatefractime}
%            %$q\leftarrow \min\{|Z|, \frac{G-\sum_{i=1}^n b_i*|w_i||\vec{w}|\cdot \vec{b})}{u_j}\}$\\
%        }{
%            $q\leftarrow \min\{|Z_j|, \lfloor\frac{G}{2d_j}-\frac{R(\vec{w}(\vec{b}))}{\mu_j}\rfloor\}$\\ \label{ln:allocatetime}
%        }
        \If{$q>0$}{
            $\mathcal{W} \leftarrow \mathcal{W} \bigcup \{j\}$~\\
            $y_j(\vec{b})\leftarrow$ The first $q$ time units in $Z_j$ \label{ln:sel}~\\
            $\mathcal{D}\leftarrow \mathcal{D}-\{A_i\in \mathcal{A}|T_i\subseteq \bigcup_{\ell:\kappa_\ell = \kappa_i}y_\ell(\vec{b})\}$~\\
        }
    }\label{ln:lpend}
    \lFor{$i\leftarrow 1$ \KwTo $n$}{
        $p_i(\vec{b})\leftarrow d_i\cdot|y_i(\vec{b})|$ \label{ln:ir}~\\
    }
%   \Return{$\vec{w}(\vec{b})=(w_1(\vec{b}),w_2(\vec{b}),...,w_n(\vec{b}))$}
    \Return{$\vec{y}(\vec{b}),~\mathcal{W},~\vec{p}(\vec{b})$}
    \caption{$\mathsf{ApproxMCS}(n, G, \vec{b}, \mathcal{A})$}
    \label{alg:assc}
  \end{algorithm}

  A partial order $\prec$ on the set $\mathcal{A}$ is used in \textbf{Algorithm~\ref{alg:assc}}, which is defined as follows. For any two users $A_i$ and $A_j$, if $\mu_i/d_i<\mu_j/d_j$ or $\mu_i/d_i=\mu_j/d_j$ but $j>i$, then we say $A_j$ \textit{suppress} $A_i$ and denote it by $A_i\prec A_j$. For any $\mathcal{A}' \subseteq \mathcal{A}$, we define $\max_{\prec}(\mathcal{A}')$ to be the user in $\mathcal{A}'$ such that there does not exist another user $\bar{A}\in \mathcal{A}'$ satisfying $\max_{\prec}(\mathcal{A}')\prec \bar{A}$. %We also define $\min_{\prec}(\mathcal{A}')$ similarly.

% \textbf{Algorithm}~\ref{alg:assc} actually contains two approximation algorithms for the MCS problem.
  \textbf{Algorithm~\ref{alg:assc}} iterates among the users and finds the schedule for them based on a greedy strategy. The algorithm, at the beginning of each iteration, selects a user $A_j$ based on the partial order $\prec$ (line~\ref{ln:greedysel}), and then computes the time units that can potentially be scheduled for $A_j$ (line~\ref{ln:optionalperiod}). The sensing time $y_j(\vec{b})$ scheduled for $A_j$ is taken as the early sub-period of the \textit{uncovered} time of $A_j$, where any time point is called uncovered if no user has been scheduled for it (line~\ref{ln:sel}). At the end of an iteration, all users whose available time periods have been covered are removed from the user set $\mathcal{D}$ upon which the algorithm iterates.
% and an early sub-period means that any time point in the sub-period is as early as possible.
  The algorithm determines the length of $y_j(\vec{b})$ based on the rule in line~\ref{ln:allocatetime}, which can be deemed as a \textit{potential function} \cite{Cormen2001} that facilitates our later quest for an approximation ratio. Also, the rule serves as a constraint to bound the total payments below $G$.

% Note that \textbf{Algorithm}~\ref{alg:assc} actually contains two approximation algorithms for the MCS problem  according to the input parameter $\beta$. When $\beta=1$, the algorithm determines the length of $w_j(\vec{b})$ simply based on the leftover budget (line~\ref{ln:allocatefractime}), and $|w_j(\vec{b})|$ can be fractional. When $\beta=0$, the selection rule of $w_j(\vec{b})$ is more tricky (line~\ref{ln:allocatetime}), and $|w_j(\vec{b})|$ is guaranteed to be integral. As performing any task for less than one unit time has no value, the fractional output of \textbf{Algorithm}~\ref{alg:assc} is only used as a benchmark and to help deriving the approximation ratio of the algorithm when $\beta=0$.

  Let us denote an iteration (from line~\ref{ln:lpbegin} to \ref{ln:lpend}) in which line~\ref{ln:sel} is executed as an \textit{effective iteration} (i.e., the concerned user is assigned a non-empty schedule). Suppose that \textbf{Algorithm~\ref{alg:assc}} has in total $h$ effective iterations. Let the user scheduled in the $i$th effective iteration be $A_{n_i}: 1 \leq i\leq h$ and let $\mathcal{W}=\{n_1,n_2,...,n_h\}$ be the \textit{index set} of these scheduled users. Let $\vec{y}_i(\vec{b})$ be the current value of vector $\vec{y}(\vec{b})$ after the $i$th effective iteration is executed, we have the following results. %The correctness of \textbf{Algorithm~\ref{alg:assc}} is proved in \textit{\textbf{Theorem}~\ref{thm:correctness}}:

  \begin{theorem}
    The output of \textbf{Algorithm}~\ref{alg:assc} is a feasible solution to the MCS problem.
    \label{thm:correctness}
  \end{theorem}

  \begin{theorem}
    Let $\vec{y}^*$ be an optimal schedule vector for the MCS problem and $\Lambda= \max_{i:d_i\leq G} (\mu_i\cdot |T_i|)$. If
%   \begin{eqnarray}
%   \max_{1\leq i\leq n} \left(\mu_i\cdot \min\{\lfloor\frac{G}{b_i}\rfloor,|T_i|\}\right) \leq \left(\frac{e-1}{2e}-\epsilon\right)R(\vec{w}^*), \label{eqn:biggest}
%   \end{eqnarray}
    \begin{eqnarray}
%     \Lambda &\leq& \left(\frac{\mathbf{e}-1}{4\mathbf{e}}-\epsilon\right)R(\vec{y}^*), \label{eqn:biggest}
      \Lambda &\leq& \left[\frac{\mathbf{e}-1}{4\mathbf{e}}-\epsilon\right]R(\vec{y}^*), \label{eqn:biggest}
    \end{eqnarray}
    for any $\epsilon\in (0,\frac{\mathbf{e}-1}{4\mathbf{e}})$, then $\vec{y}$ is a $\frac{4}{3}\epsilon$ approximation to MCS.
    \label{thm:conditionalar}
  \end{theorem}
  Note that an extra condition (\ref{eqn:biggest}) needs to be satisfied in \textit{\textbf{Theorem}~\ref{thm:conditionalar}}, hence the approximation ratio is conditional. We shall handle this issue using a randomized mechanism design method in the next section.

\section{Offline Mechanisms for MCS} \label{sec:offline}
  As users' strategic behaviors are not considered in \textbf{Algorithm~\ref{alg:assc}}, the payments to all users made there are directly determined by the declared sensing cost and the length of the time periods scheduled for them. Unfortunately, this method can be non-truthful: a user may lie about his/her private information to manipulate the length of his/her scheduled time, hence to gain a higher utility. Characterizations of truthful mechanisms exist in the literature (e.g., \cite{Myerson1981,Archer2001}), but these characterizations are only for single parameter mechanisms, while in our problem any user $A_i$ has three parameters, namely $d_i$, $s_i$ and $e_i$. Therefore, we hereby design a novel truthful mechanism for the MCS problem under the offline setting, as shown by \textbf{Algorithm~\ref{alg:mch}}. A sub-routine used by \textbf{Algorithm~\ref{alg:mch}} to compute payments to users is shown in \textbf{Algorithm~\ref{alg:payment}}.
  \begin{algorithm}[h!]
%    \KwIn{$G, \vec{b}, \mathcal{A}, T$}
%    %
%    \KwOut{$S, w$}
    %
    Generate a random number $o$ from the uniform distribution on the interval [0, 1]~\\
    \If{$o\leq \frac{1}{2}$}{
        %Call $\mathsf{ApproxMCS}(n,G, \vec{b}, \mathcal{A})$ to get the sensing schedule $\vec{y}$;\\
        $\left(\vec{y}(\vec{b}),\mathcal{W}\right) \leftarrow \mathsf{ApproxMCS}(n,G, \vec{b}, \mathcal{A})$ \label{ln:callassc}~\\
        %Call $\mathsf{CalPayment}(n,G, \vec{b}, \mathcal{A}, \mathcal{T},\mathcal{W}, \vec{y}, n_i)$ to calculate the payment to each $A_{n_i}$ and pay $A_{n_i}$ at the end of $T_{n_i}(1\leq i\leq h)$; pay other users zero;\\
        \ForAll{$n_i \in \mathcal{W}$} {
            $p_{n_i}(\vec{b}) \leftarrow \mathsf{CalPayment}(n, G, \vec{b}, \mathcal{A},\mathcal{W}, \vec{y}, n_i)$ \label{ln:callpay}~\\
        }
        \lForAll{$i \not\in \mathcal{W}$}{
            $p_i(\vec{b}) \leftarrow 0$ \label{ln:endcal}
        }
%       Pay $p_{n_i}(\vec{b})$ to $A_{n_i}\!: n_i \in \mathcal{W}$ at the end of $T_{n_i}: 1\leq i\leq h$, and pay other users zero
    }
    \Else{
        \lForAll{$i \in \mathcal{A}$}{
            $p_i(\vec{b}) \leftarrow 0$ \label{ln:bstart}~\\
        }
        $j\leftarrow \arg\max_{i:d_i\leq G} \mu_i$;~~~$\mathcal{W} \leftarrow \{j\}$~\\
        $y_j(\vec{b}) \leftarrow$ an arbitrary time unit in $T_j$;~~~$p_j(\vec{b}) \leftarrow G$ \label{ln:bend}
%       Find any sub-period in $T_j$ with a length of one time unit as $A_j$'s sensing schedule \\
%       Pay $A_j$ the amount $G$ and pay others zero \label{ln:bend} \\
%       $j\leftarrow \arg\max_{1\leq i \leq n}\left(\mu_i\cdot \min\{\lfloor\frac{G}{b_i}\rfloor,|T_i|\}\right)$\\
%       Find any sub-period in $T_j$ with $\min\{|T_j|,\lfloor\frac{G}{b_j}\rfloor\}$ time slots as $A_j$'s sensing schedule\\
%       Pay $A_j$ the amount $G$ and pay others zero.\\
    }
    Pay $p_i(\vec{b})$ to $A_i\!: \forall i \in \mathcal{W}$ at the end of $T_i$ if $A_i$ successfully completes his/her sensing task during $y_i(\vec{b})$ \label{ln:postpaid}\\
    \Return{$\vec{y}(\vec{b}),~\vec{p}(\vec{b})$}
%   \Return{$\langle S, w\rangle$}
    \caption{Truthful Offline Mechanism for MCS}
    \label{alg:mch}
  \end{algorithm}

  \begin{algorithm}[htb]
%    \KwIn{$n_1,n_2,...,n_h,\vec{y},\vec{b},G, \mathcal{A}, \mathcal{T}$}
%    %
%    \KwOut{${p}_{n_i}(\vec{b})$}
    \lFor{$j\leftarrow 1$ \KwTo $n$}{
        $t_j\leftarrow \emptyset$ \label{ln:calpaybeforestart}~\\
    }
    \lFor{$j\leftarrow 1$ \KwTo $i-1$}{
        $t_{n_j}\leftarrow y_{n_j}(\vec{b})$ \label{ln:calpaybeforeend}~\\
    }
    $\mathcal{S}\leftarrow \{j| 1\leq j\leq n, \frac{\mu_j}{d_j} \prec \frac{\mu_{n_i}}{d_{n_i}}\}$ \label{ln:setS}~\\
    $p_{n_i}(\vec{b})\leftarrow d_{n_i}\cdot |y_{n_i}(\vec{b})|;~~~k\leftarrow n_i;~~~\theta\leftarrow |\mathcal{S}|$ \\
    \While{$\theta\geq 0$}{ \label{ln:repeatstart}
        $\mathcal{Z}_1\leftarrow T_{n_i}-\bigcup_{\ell:\kappa_{n_i}=\kappa_{\ell}}t_{\ell}$;~$\gamma_1\leftarrow \frac{\mu_{n_i}}{\mu_{k}}\cdot d_{k}$;~$\gamma_2\leftarrow \frac{\mu_{n_i}G}{2R(\vec{t})}$ \label{ln:fracpaystart}~\\
        \If{$\mathcal{S}\neq \emptyset$}{
            Find $j$ such that $A_j=\max_{\prec}(\mathcal{S})$~\\
            $\gamma_2 \leftarrow \min\left\{\frac{\mu_{n_i}}{\mu_{{j}}}\cdot d_{{j}},\gamma_2\right\}$~\\
            %$j\leftarrow \arg\min_{\ell\in S} \frac{\mu_{\ell}}{d_{\ell}}$\\
            }
        \uIf{$|\mathcal{Z}_1|> 0\bigwedge \gamma_2\geq \gamma_1$}{
            $p_{n_i}(\vec{b})=p_{n_i}(\vec{b})\!+\!\int_{\gamma_1}^{\gamma_2} \min\left\{|\mathcal{Z}_1|,\left\lfloor\frac{G}{2v}-\frac{R(\vec{t})}{\mu_{n_i}}\right\rfloor\right\} \mathrm{d}v$ \label{ln:calfracpayment}
        }\lElse{$\mathbf{break}$} \label{ln:fracpayend}~\\
        \If{$\mathcal{S}\neq\emptyset$}{ \label{ln:subsitutestart}
        $\mathcal{Z}_2\leftarrow T_{j}-\bigcup_{\ell:\kappa_{j}=\kappa_{\ell}}t_{\ell}$~\\
        $q\leftarrow \min\left\{|\mathcal{Z}_2|, \left\lfloor\frac{G}{2d_{j}}-\frac{R(\vec{t})}{\mu_{j}}\right\rfloor\right\}$~\\
        \If{$q>0$}{
            $t_{j}\leftarrow$ The first $q$ time units in $Z_2$~\\
        }
        \lIf{$q<|\mathcal{Z}_2|$}{$\mathbf{break}$}~\\
        $k\leftarrow j;~~~\theta\leftarrow |\mathcal{S}|;~~~\mathcal{S}\leftarrow \mathcal{S}\backslash \{j\}$ \\ \label{ln:subsituteend}
        }
        $\theta\leftarrow \theta-1$ \label{ln:repeatend}
    }
%    \If{$S=\emptyset$}{
%        $\mathcal{Z}_1\leftarrow T_{n_i}-\bigcup_{l:\kappa_{n_i}=\kappa_{l}}z_{l}$\\
%        $\mathit{\gamma_1}\leftarrow \frac{\mu_{n_i}}{\mu_{k}}\cdot d_{k}$\\
%        $\mathit{\gamma_2}\leftarrow \min\{\frac{\mu_{n_i}}{\mu_{{j}}}\cdot d_{{j}},\frac{\mu_{n_i}G}{2R(\vec{z})}\}$\\
%    }
    %}
    \Return{$p_{n_i}(\vec{b})$}
    \caption{$\mathsf{CalPayment}(n,G, \vec{b}, \mathcal{A},\mathcal{W}, \vec{y},n_i)$} \label{alg:payment}
  \end{algorithm}

  \textbf{Algorithm~\ref{alg:mch}} is apparently a randomized mechanism. With probability one half, the algorithm calls \textbf{Algorithm~\ref{alg:assc}} to get a feasible schedule for each user (line~\ref{ln:callassc}). However, instead of using the simple payment rule in \textbf{Algorithm}~\ref{alg:assc}, \textbf{Algorithm~\ref{alg:mch}} replaces it with a more complicated method shown by \textbf{Algorithm~\ref{alg:payment}} to calculate the payments (line~\ref{ln:callpay}); otherwise \textbf{Algorithm~\ref{alg:mch}} runs lines~\ref{ln:bstart}-\ref{ln:bend} and selects a user whose sensing cost per unit time is no more than the budget and whose sensing value per unit time is maximized. Then the selected user %is scheduled for one time unit and
  is paid the amount $G$, while others are paid zero. The payments are made to the users using a \textit{post-paid} scheme, i.e., a payment is made instantly at the end of a user's claimed available time period only if he/she has successfully performed the sensing task during the whole time period scheduled for him/her (line~\ref{ln:postpaid}). To understand the payment calculation in \textbf{Algorithm~\ref{alg:payment}}, we introduce \textit{\textbf{Lemma}~\ref{lma:largevalueisless}}, \textit{\textbf{Lemma}~\ref{lma:smallintervalisless}} and \textit{\textbf{Theorem}~\ref{thm:payment}}, which are also useful for characterizing truthfulness under our multi-parameter environment. % \textit{\textbf{Lemma}~\ref{lma:largevalueisless}}, \textit{\textbf{Lemma}~\ref{lma:smallintervalisless}} and \textit{\textbf{Theorem}~\ref{thm:payment}}, which actually provide a characterization of truthfulness (in terms of sufficiency) under our multi-parameter environment.
% The essence of \textbf{Algorithm}~\ref{alg:payment} is illustrated by Lemma~\ref{lma:payment}:
  \begin{lemma}
    For any user $A_i$ and his/her two bids $b_i=(d_i, s_i,e_i)$ and $b'_i=(d_i',s_i,e_i)$, if $d_i'\geq d_i$, then $|y_i(b'_i,b_{-i})|\leq |y_i(b_i,b_{-i})|$.
    \label{lma:largevalueisless}
  \end{lemma}
  \begin{lemma}
    For any user $A_i$ and his/her two bids $b_i=(d_i, s_i,e_i)$ and $b'_i=(d_i,s'_i,e'_i)$, if $[s'_i,e'_i]\subseteq [s_i,e_i]$, then $|y_i(b'_i,b_{-i})|\leq |y_i(b_i,b_{-i})|$.
    \label{lma:smallintervalisless}
  \end{lemma}
  \begin{theorem}
    For any user $A_i$ with the bid $b_i=(d_i,s_i,e_i)$ and any $b_{-i}$, $\int_0^\infty |y_i\left((v,s_i,e_i),b_{-i}\right)|\mathrm{d}v<+\infty$ and the payment to $A_i$ computed by \textbf{Algorithm~\ref{alg:payment}} is
      \begin{equation}
        p_i(b_i,b_{-i}) = d_i|y_i(b_i,b_{-i})| + \!\int_{d_i}^\infty \!\!|y_i\left((v,s_i,e_i),b_{-i}\right)|\mathrm{d}v.  \label{eqn:payformulation}
      \end{equation}
    \label{thm:payment}
  \end{theorem}

  \begin{IEEEproof}
    For $i\not \in \{n_1,n_2,...,n_h\}$, we have $p_i(b_i,b_{-i})=0$ and $|y_i(b_i,b_{-i})|=0$. According to \textit{\textbf{Lemma}~\ref{lma:largevalueisless}}, for any $v>d_i$, we have $|y_i((v,s_i,e_i),b_{-i})|\leq|y_i(b_i,b_{-i})|$, hence $|y_i((v,s_i,e_i),b_{-i})|=0$. So (\ref{eqn:payformulation}) holds in this case.

    Now we analyze \textbf{Algorithm~\ref{alg:payment}} in details for the case of $i\in \{n_1,n_2,...,n_h\}$, and we write $i$ as $n_i$ in this case. In \textbf{Algorithm~\ref{alg:payment}}, we first initialize the vector $\vec{t}$ to record the user schedules that are decided before $A_{n_i}$, as shown in lines~\ref{ln:calpaybeforestart}-\ref{ln:calpaybeforeend}. Then we use $\mathcal{S}$ to denote the indices of the users that are suppressed by $A_{n_i}$ according to the partial order $\prec$ (line~\ref{ln:setS}). Note that if $A_{n_i}$ bids $d'_{n_i}> d_{n_i}$, the uncovered available time period of $A_{n_i}$ (i.e., $\mathcal{Z}_1$) may change, because we have $\mu_{n_i}/d'_{n_i}<\mu_{n_i}/d_{n_i}$ and some users %$A_{j} (\mu_j/d_j\prec \mu_i/d_i)$
    originally suppressed by $A_{n_i}$ may hence get scheduled before $A_{n_i}$. Consequently, we calculate the schedule for other users (recorded in $\vec{t}$) when $d_{n_i}$ increases, and divide the interval $[d_{n_i},+\infty)$ into some sub-intervals, such that the uncovered available time period of $A_{n_i}$ remains the same when $A_{n_i}$ changes $d_{n_i}$ within each of these sub-intervals, as shown in lines~\ref{ln:repeatstart}-\ref{ln:repeatend}. More specifically, the algorithm, at the beginning of each iteration, picks a user's index $j$ from $\mathcal{S}$ such that $A_j$ is maximal with respect to the partial order $\prec$, then it identifies a sub-interval $\left(\left.\frac{\mu_{n_i}}{\mu_{{k}}}\cdot d_{{k}},\frac{\mu_{n_i}}{\mu_{{j}}}\cdot d_{{j}}\right]\right.$ for $d_{n_i}$ (indicated by $\gamma_1$ and $\gamma_2$), where $k$ is the index of the last picked user from $\mathcal{S}$ (line~\ref{ln:subsituteend}, but initially set as $n_i$). As $A_{n_i}$ remains scheduled before $A_j$ when $d_{n_i}$ varies within this sub-interval, we calculate the partial payment in line~\ref{ln:calfracpayment} based on the current schedule. %lines~\ref{ln:fracpaystart}-\ref{ln:fracpayend} based on the scheduling rule in \textbf{Algorithm~\ref{alg:assc}}.
    When $d_{n_i}$ gets bigger than $\frac{\mu_{n_i}}{\mu_{{j}}}\cdot d_{{j}}$, then $A_j$ will be scheduled before $A_{n_i}$, and we calculate $A_{j}$'s schedule in this case by lines~\ref{ln:subsitutestart}-\ref{ln:subsituteend}. With this adjusted schedule, the algorithm goes into the next iteration to further accumulate the partial payment in a different sub-interval.
    %After all the users in $\mathcal{S}$ are picked out, $d_{n_i}$ has increased to a value that makes $A_{n_i}$ to be scheduled at the last place. In such a case, we calculate the partial payment to $A_{n_i}$ similarly.
    At the end, the algorithm goes through all the possible sub-intervals in $[d_{n_i}, \infty)$, so the payment is exactly calculated as the right-hand side of (\ref{eqn:payformulation}). Hence the theorem follows.
  \end{IEEEproof}

%\begin{algorithm}[htb]
%%    \KwIn{$G, \vec{b}, \mathcal{A}, T$}
%%    %
%%    \KwOut{$S, w$}
%    %
%    Generate a random number $o$ from the uniform distribution on the interval [0, 1];\\
%    \eIf{$o\leq \frac{1}{2}$}{
%        Run \textbf{Algorithm}~\ref{alg:assc} with $\beta=0$ to get the sensing schedule $\vec{y}$;\label{ln:astart}\\
%        Run the payment algorithm to calculate the payment to each $A_{n_i}(1\leq i\leq h)$ and pay others zero;\label{ln:aend}\\
%    }{
%        $j\leftarrow \arg\max_{i:b_i\leq G} \mu_i$\label{ln:bstart}\\
%        Find any sub-period in $T_j$ with one time slot as $A_j$'s sensing schedule\\
%        Pay $A_j$ the amount $G$ and pay others zero.\label{ln:bend}\\
%%        $j\leftarrow \arg\max_{1\leq i \leq n}\left(\mu_i\cdot \min\{\lfloor\frac{G}{b_i}\rfloor,|T_i|\}\right)$\\
%%        Find any sub-period in $T_j$ with $\min\{|T_j|,\lfloor\frac{G}{b_j}\rfloor\}$ time slots as $A_j$'s sensing schedule\\
%%        Pay $A_j$ the amount $G$ and pay others zero.\\
%    }
%%    \Return{$\langle S, w\rangle$}
%    \caption{A truthful offline scheduling mechanism} \label{alg:mch}
%\end{algorithm}

%\begin{lemma}
%$\sum_{1\leq i\leq h} b_{n_i}\cdot |y_{n_i}| \leq \frac{G}{2}$
%\end{lemma}

%$Y\preceq Z$ if and only if for any $\langle j,w_j\rangle\in Y$ there exists $\langle j,w_j'\rangle\in Z$ such that $w_j\subseteq w_j'$. Let $Z\thicksim Y=\{\langle j, w_j'\backslash w_j\rangle | \langle j, w_j\rangle \in Y\wedge \langle j, w_j'\rangle \in Z\}$. Then we have
%
%\begin{lemma}
%If $Y\preceq Z$, then
%\end{lemma}

  In \textit{\textbf{Lemma}~\ref{lma:biddingbound}} and \textit{\textbf{Theorem}~\ref{thm:correctmch}}, we prove that the payment calculated by \textbf{Algorithm~\ref{alg:payment}} is no more than the budget $G$ and \textbf{Algorithm~\ref{alg:mch}} provides a feasible solution satisfying IR to the MCS problem.
  %hence the mechanism shown in \textbf{Algorithm}~\ref{alg:mch} provides a feasible solution to the MCS problem.
  %
  \begin{lemma}
    For any user $A_j$ and his/her bid $(d_{j},s_{j},e_{j})$, if $j\in \{n_1,n_2,...,n_h\}$, then we have $d_{j}\leq \mu_{j}\cdot G/R(\vec{y})$.
    \label{lma:biddingbound}
  \end{lemma}
  \begin{theorem}
    The mechanism shown in \textbf{Algorithm~\ref{alg:mch}} provides a feasible solution that satisfies IR to the MCS problem.
    \label{thm:correctmch}
  \end{theorem}

  More importantly, using the results stated in \textit{\textbf{Lemma}~\ref{lma:largevalueisless}}, \textit{\textbf{Lemma}~\ref{lma:smallintervalisless}} and \textit{\textbf{Theorem}~\ref{thm:payment}}, we can now prove the truthfulness of the mechanism by \textit{\textbf{Theorem}~\ref{thm:truthful}}. The rationale lies in the difference between any user $A_i$'s payment and his/her true sensing cost: regardless of how other users may bid, this difference is always maximized if $A_i$ bids truthfully and hence truth-telling is a dominant strategy for $A_i$.

  \begin{theorem}
    The scheduling mechanism shown in \textbf{Algorithm~\ref{alg:mch}} is truthful.% and individually rational.
    \label{thm:truthful}
  \end{theorem}
  \begin{IEEEproof}
    We first prove that lines~\ref{ln:callassc}-\ref{ln:endcal} is truthful. Suppose that there exists a user $A_i$ whose truthful bid is $\hat{b}_i=(\hat{d}_i,\hat{s}_i,\hat{e}_i)$, but he/she can get a higher utility by bidding $b_i=(d_i,s_i,e_i)\neq \hat{b}_i$ for some $b_{-i}$. If $y_i(b_i,b_{-i})\not\subseteq [\hat{s}_i,\hat{e}_i]$, $A_i$ gets zero payment because %he/she cannot perform sensing at his/her unavailable time and
    the mechanism requires $A_i$ to complete sensing during $y_i(b_i,b_{-i})$  to get paid. If $e_i\not\in [\hat{s}_i,\hat{e}_i]$, $A_i$ again gets zero payment because the mechanism employs a post-paid scheme and $A_i$ cannot get paid when he/she is unavailable. Therefore, we must have  $y_i(b_i,b_{-i})\subseteq [\hat{s}_i,\hat{e}_i]$ and $e_i\in [\hat{s}_i,\hat{e}_i]$. Now if $s_i\geq \hat{s}_i$, we have $[s_i,e_i]\subseteq [\hat{s}_i,\hat{e}_i]$; so using \textit{\textbf{Lemma}~\ref{lma:smallintervalisless}} we get
    \begin{eqnarray}
      |y_i\left((d_i,s_i,e_i),b_{-i}\right)| &\leq& |y_i\left((d_i,\hat{s}_i,\hat{e}_i),b_{-i}\right)|.
      \label{eqn:key1}
    \end{eqnarray}
    Otherwise if $s_i < \hat{s}_i$, we know that the period $[s_i, \hat{s}_i]$ must have been covered before deciding the schedule of $A_i$ based on his/her bidding $b_i$, because otherwise the algorithm will allocate time in $[s_i, \hat{s}_i]$ to $A_i$ according to line~\ref{ln:sel} of \textbf{Algorithm~\ref{alg:assc}}, which contradicts $y_i(b_i,b_{-i})\subseteq [\hat{s}_i,\hat{e}_i]$. Hence we know $y_i\left((d_i,s_i,e_i),b_{-i}\right)=y_i\left((d_i,\hat{s}_i,e_i),b_{-i}\right)$. As $[\hat{s}_i,e_i]\subseteq [\hat{s}_i,\hat{e}_i]$, (\ref{eqn:key1}) also holds by using \textit{\textbf{Lemma}~\ref{lma:smallintervalisless}}.

    For $A_i$'s any bid $(d'_i,s'_i,e'_i)$, let $f_i((d'_i,s'_i,e'_i),b_{-i})$ denote the uncovered time in $[s'_i,e'_i]$ when the algorithm allocates time to $A_i$ based on a bid vector $((d'_i,s'_i,e'_i),b_{-i})$. The above reasoning actually reveals that $f_i((d_i,s_i,e_i),b_{-i})\subseteq f_i((d_i,\hat{s}_i,\hat{e}_i),b_{-i})$. According to the mechanism, for any $v\geq d_i$ we have $f_i((v,s_i,e_i),b_{-i})\subseteq f_i((d_i,s_i,e_i),b_{-i})$, $f_i((v,\hat{s}_i,\hat{e}_i),b_{-i})\subseteq f_i((d_i,\hat{s}_i,\hat{e}_i),b_{-i})$ and
    \begin{eqnarray}
        [f_i((d_i,\hat{s}_i,\hat{e}_i),b_{-i})\backslash f_i((v,\hat{s}_i,\hat{e}_i),b_{-i})]\cap f_i((d_i,s_i,{e}_i),b_{-i})\nonumber\\
        \subseteq f_i((d_i,{s}_i,{e}_i),b_{-i})\backslash f_i((v,{s}_i,{e}_i),b_{-i}), \nonumber
    \end{eqnarray}
%    \begin{eqnarray}
%      &&f_i((d_i,s_i,e_i),b_{-i})-f_i((v,s_i,e_i),b_{-i}) \nonumber\\
%      &\subseteq&f_i((d_i,\hat{s}_i,\hat{e}_i),b_{-i})-f_i((v,\hat{s}_i,\hat{e}_i),b_{-i}), \nonumber
%    \end{eqnarray}
    %
    which yield $f_i((v,s_i,e_i),b_{-i})\subseteq f_i((v,\hat{s}_i,\hat{e}_i),b_{-i})$ and
    %
%    The above reasoning reveals that the uncovered period of $[s_i,e_i]$ is a subset of $[\hat{s}_i,\hat{e}_i]$ when the algorithm allocates time for $A_i$ with the bid $(d_i,s_i,e_i)$. Since the uncovered time period in $[s_i,e_i]$ can only diminish if $A_i$ bids $(v,s_i,e_i)$ where $v\geq d_i$, we have
    %
    \begin{eqnarray}
      |y_i\left((v,s_i,e_i),b_{-i}\right)| &\leq& |y_i\left(((v,\hat{s}_i,\hat{e}_i),b_{-i}\right)|.
      \label{eqn:key2}
    \end{eqnarray}
%   This implies that $e'_i\geq s_i$ and $w_i((d'_i,s'_i,e'_i),b_{-i})\subseteq [s_i,e'_i]$. According to (i), we have
%   \begin{eqnarray}
%   w_i((d'_i,s'_i,e'_i),b_{-i})= w_i((d'_i,s_i, e'_i),b_{-i})
%   \end{eqnarray}
%   According to (ii), we know that $w_i((d'_i,s_i, e'_i),b_{-i})\subseteq w_i((d'_i,s_i,e_i),b_{-i})$. Hence, we have
%   \begin{eqnarray}
%   w_i((d'_i,s'_i,e'_i),b_{-i})\subseteq w_i((d'_i,s_i,e_i),b_{-i})
%   \label{eqn:key}
%   \end{eqnarray}
    Since the user gets more utility by bidding $b_i$ then by bidding $\hat{b}_i$, we know that:
    \begin{eqnarray}
      &&p_i((d_i,s_i,e_i),b_{-i})- \hat{d}_i\cdot |y_i((d_i,s_i,e_i),b_{-i})|\nonumber\\
      &>& p_i((\hat{d}_i,\hat{s}_i,\hat{e}_i),b_{-i})- \hat{d}_i\cdot |y_i((\hat{d}_i,\hat{s}_i,\hat{e}_i),b_{-i})|. \nonumber
    \end{eqnarray}
    Combing this with \textit{\textbf{Theorem}~\ref{thm:payment}} gives us
    \begin{eqnarray}
      &&(d_i- \hat{d}_i)\cdot |y_i((d_i,s_i,e_i),b_{-i})| \nonumber\\
      &>& \int_{\hat{d}_i}^\infty |y_i((v,\hat{s}_i,\hat{e}_i),b_{-i})|\mathrm{d}v-\int_{d_i}^\infty |y_i((v,s_i,e_i),b_{-i})|\mathrm{d}v. \nonumber
    \end{eqnarray}

    Case 1: $d_i\geq \hat{d}_i$, using (\ref{eqn:key2}) and \textit{\textbf{Lemma}~\ref{lma:largevalueisless}} we get
    %
%   According to equation~(\ref{eqn:key}), the period $T'_i\backslash T_i$ must have been covered when deciding the schedule of $A_i$ for his bidding $(d'_i,s'_i,e'_i)$, otherwise the algorithm will allocate time in $T'_i\backslash T_i$ to $A_i$, which is a contradiction to equation~(\ref{eqn:key}). In other words, this means that the uncovered sub-period of $T'_i$ at that time is a sub-period of $T_i$. Therefore, for any $v\geq d'_i$, we have
%   \begin{eqnarray}
%   |w_i((v,T'_i),b_{-i})|\leq |w_i((v,T_i),b_{-i})|
%   \end{eqnarray}
%
%   Now, if $d'_i\geq d_i$, then using equation~(\ref{eqn:key}) and (iii) we know that,
    %
    \begin{eqnarray}
      &&(d_i- \hat{d}_i)\cdot |y_i((d_i,s_i,e_i),b_{-i})| \nonumber \\
      &>& \int_{\hat{d}_i}^{d_i} |y_i((v,\hat{s}_i,\hat{e}_i),b_{-i})|\mathrm{d}v \nonumber\\
      &\geq& (d_i- \hat{d}_i)\cdot |y_i((d_i,\hat{s}_i,\hat{e}_i),b_{-i})|. \nonumber
    \end{eqnarray}
    If $d_i=\hat{d}_i$, then we get $0>0$, a contradiction. If $d_i>\hat{d}_i$, then we get $|y_i((d_i,s_i,e_i),b_{-i})|>|y_i((d_i,\hat{s}_i,\hat{e}_i),b_{-i})|$, which contradicts (\ref{eqn:key1}).

    Case 2: $d_i<\hat{d}_i$, using (\ref{eqn:key2}) and \textit{\textbf{Lemma}~\ref{lma:largevalueisless}} we get:
%   On the other side, if $d_i>d'_i$, using equation~(\ref{eqn:key}) and (iii) we know that, for any $v\geq d_i$, we get
%   \begin{eqnarray}
%   |w_i((v,T'_i),b_{-i})|\leq |w_i((v,T_i),b_{-i})| \nonumber
%   \end{eqnarray}
%   Hence,
    \begin{eqnarray}
      &&(d_i- \hat{d}_i)\cdot |y_i((d_i,s_i,e_i),b_{-i})| \nonumber \\
      &>& -\int_{d_i}^{\hat{d}_i} |y_i((v,s_i,e_i),b_{-i})|\mathrm{d}v \nonumber\\
      &\geq& (d_i- \hat{d}_i)\cdot |y_i((d_i,s_i,e_i),b_{-i})|, \nonumber
    \end{eqnarray}
    hence $|y_i((d_i,s_i,e_i),b_{-i})| <|y_i((d_i,s_i,e_i),b_{-i})|$, also a contradiction.

    The above reasoning has shown that lines~\ref{ln:callassc}-\ref{ln:endcal} is truthful. Now we prove that lines~\ref{ln:bstart}-\ref{ln:bend} is truthful. If a user $A_i$ gets a non-empty schedule by bidding truthfully, then $A_i$ clearly cannot benefit from lying. Now suppose that $A_i$ gets an empty schedule (hence the utility 0) by bidding truthfully. If $\hat{d}_i \leq G$, then $A_i$ cannot increase his/her utility by lying, because he/she will anyway get an empty schedule regardless of his/her bid. If $\hat{d}_i>G$, then the only way that may allow $A_i$ to get a non-empty schedule is to bid some $(d_i,s_i,e_i)$ with $d_i\leq G$. However, in that case $A_i$'s utility is $G-\hat{d}_i<0$. Therefore, $A_i$ is better off bidding truthfully. From the above reasoning, we know that \textbf{Algorithm~\ref{alg:mch}} is a randomization of two truthful mechanisms, and is hence truthful.
  \end{IEEEproof}

  Finally, based on the approximation ratio of \textbf{Algorithm~\ref{alg:assc}}, we can prove that \textbf{Algorithm~\ref{alg:mch}} has an $\mathcal{O}(1)$ approximation ratio, as shown by \textit{\textbf{Theorem}~\ref{thm:mchar}}. We also analyze the time complexity of \textbf{Algorithm}~\ref{alg:mch} by \textit{\textbf{Theorem}~\ref{thm:timcom}}.
  \begin{theorem}
    The mechanism shown in \textbf{Algorithm~\ref{alg:mch}} has an approximation ratio of $\mathcal{O}(1)$.
    \label{thm:mchar}
  \end{theorem}
  \begin{theorem}
    The worst-case time complexity of \textbf{Algorithm~\ref{alg:mch}} is $\mathcal{O}(n^2)$.
    \label{thm:timcom}
  \end{theorem}
  Obviously, the major time complexity results from calling \textbf{Algorithm~\ref{alg:payment}} for at most $n$ times.

%
%\begin{proof}
%For any user $A_i$ and his two bids $b_i=(d_i, T_i)$ and $b'_i=(d_i,T'_i)$, if $w((d_i,T'_i),b_{-i})\subseteq w((d_i,T_i),b_{-i})$, we know that $w((d_i,T'_i),b_{-i})\subseteq T'_i\cap T_i$. This means that
%\end{proof}

%\begin{algorithm}[htb]
%    \KwIn{}
%    %
%    \KwOut{}
%    %
%    $C\leftarrow [n]; T\leftarrow \emptyset$ \\
%    $\forall i: w_i\leftarrow \emptyset$\\
%    \While {$B>0$ and $C\neq \emptyset$} {
%        $j\leftarrow \arg \max_{i\in C} \frac{v_i}{b_i}$ \\
%%        $p\leftarrow p+1$\\
%        \eIf{$B-|T_j\backslash T|\cdot b_j>0$}{
%            $w_j=T_j\backslash T$
%        }{
%            let $w_j$ be any subset of $T_j\backslash T$ such that $|w_j|\cdot b_j=B$
%        }
%        $S\leftarrow S\cup \{\langle j, w_j\rangle\}$\\
%        $B\leftarrow \max\{0, B-|T_j\backslash T|\cdot b_j\}$ \\
%        $T\leftarrow T\cup T_j$\\
%        $C\leftarrow C\backslash \{i|T_i\subseteq T\}$\\
%    }
%    \caption{Finding an approximate MSB} \label{alg:MSB}
%\end{algorithm}

%\begin{table}[!t]%[htbp]
%\caption{Symbols and Notations} \vspace{-.5ex}
%%\resizebox{1.1\columnwidth}{!}{
%\begin{tabular}
%{|p{37pt}|p{180pt}|}
%\hline
%Notation&
%\quad\quad\quad\quad\quad\quad\quad\quad~~ Description \\
%\hline
%$G$&
%The graph representing a WSN \\
%\hline
%\end{tabular}
%%}
%\vspace{-1.5ex}
%\label{tab1}
%\end{table}

\section{Online Mechanisms for MCS} \label{sec:online}
%
%\begin{algorithm}[htb]
%    \lFor{$i\leftarrow 1$ \KwTo $n$}{
%        $z_i\leftarrow \emptyset;p_i\leftarrow 0$\\
%    }
%    $\lambda\leftarrow \max_{1\leq i\leq \lfloor \frac{n}{\mathbf{e}}\rfloor}\{\mu_i\cdot \min(|T_i|,\lfloor \frac{G}{b_i}\rfloor)\}$\\
%    \For{$i\leftarrow \lfloor \frac{n}{\mathbf{e}}\rfloor+1$ \KwTo $n$}{
%        \If{$\mu_i\cdot \min(|T_i|,\lfloor \frac{G}{b_i}\rfloor)\geq \lambda$}{
%            $z_i\leftarrow$ any sub-period of $T_i$ with $\min\{|T_i|,\lfloor \frac{G}{b_i} \rfloor\}$ time slots\\
%            $p_i\leftarrow G$\\
%        }
%    }
%    \Return{$(\vec{z},\vec{p})$}
%    \caption{A naive online scheduling mechanism} \label{alg:secra}
%\end{algorithm}
  In this section, we study incentive mechanisms for the MCS problem under the online setting, where the users come in random orders and the schedule/payment for each user has to be decided upon his/her arrival.
  % one by one. We will design truthful online mechanisms with respect to the user's sensing costs, i.e.:
  We assume in this case that any user $A_i$ would only lie about his/her sensing cost $\hat{d}_i$, and we will design truthful mechanisms such that reporting his/her real cost is a dominant strategy of $A_i$.
  % maximize $A_i$'s utility when he reports his real $d_i$.
  The problem of handling users' strategic bidding on their available time periods under the online setting is left for future work.
%
%
%we assume that any user $A_i$ would bid truthfully on their available sensing time period $T_i$ but may lie about the sensing cost per unit time $d_i$; and our semi-truthful mechanisms will maximize $A_i$'s utility when he bids truthfully on $d_i$.

%We first propose a simple online mechanism (\textbf{Algorithm}~\ref{alg:secra}) for the MCS problem, whose idea originates from the secretary algorithm~\cite{Dynkin1963}. In \textbf{Algorithm}~\ref{alg:secra}, we assign empty schedule to the first arrived $\lfloor \frac{n}{\mathbf{e}}\rfloor$ users and use these users' bids as a threshold for the leftover users. More specifically, we find a user from the first come $\lfloor \frac{n}{\mathbf{e}}\rfloor$ users whose sensing cost per unit time is no more than the budget and whose sensing value per unit time is the maximum (denoted by $\alpha$), then we selects a first seen leftover user whose sensing cost per unit time is no less than $\alpha$ and pay him $G$. The truthfulness and competitive ratio of \textbf{Algorithm}~\ref{alg:secra} are given in \textit{\textbf{Theorem}~\ref{thm:secrtruthful}} and \textit{\textbf{Theorem}~\ref{thm:onlmchpartb}}, respectively:

  We first propose a simple online mechanism in \textbf{Algorithm~\ref{alg:secra}} for the MCS problem, whose idea originates from the \textit{secretary algorithm}~\cite{Dynkin1963}.
  \begin{algorithm}[htb]
    $\alpha\leftarrow 0$;~~~$j\leftarrow 0$ \\
    \SetKwBlock{Upon}{upon}{end}
    \Upon($A_i$'s arrival){
%   \While {a user with id $i$ comes} {
        $j\leftarrow j+1$\\
        \eIf{$j\leq \lfloor \frac{n}{\mathbf{e}}\rfloor$}{
            $y_i\leftarrow \emptyset$;~~~$p_i\leftarrow 0$ \\ \label{ln:rejectfirst}
            \lIf{$d_i\leq G$}{
                $\alpha\leftarrow \max\{\alpha,\mu_i\}$ \label{ln:calmaximumalpha}
                %$\alpha\leftarrow \max\{\alpha,\mu_i\cdot |T_i|\}$
            }
        }
        {
            \eIf{$\mu_i \geq \alpha \bigwedge d_i\leq G \bigwedge G>0$}{ \label{ln:secralessg}
                $y_i\leftarrow$ an arbitrary time unit in $T_i$ \\
%               $y_i\leftarrow$ any sub-period of $T_i$ with $\min\{|T_i|,\lfloor \frac{G}{b_i} \rfloor\}$ time slots\\
                $p_i\leftarrow G$;~~~$G\leftarrow 0$ \\
            }
            {
                $y_i\leftarrow \emptyset$;~~~$p_i\leftarrow 0$ \label{ln:secralessgend}
            }
        }
    }
    \Return{$(\vec{y},\vec{p})$}
    \caption{A Deterministic Online Mechanism} \label{alg:secra}
  \end{algorithm}
  In lines~\ref{ln:rejectfirst}-\ref{ln:calmaximumalpha} of \textbf{Algorithm}~\ref{alg:secra}, we assign empty schedules to the first arrived $\lfloor \frac{n}{\mathbf{e}}\rfloor$ users, and find one of them whose sensing cost per unit time is no more than the budget and whose sensing value per unit time is the maximum denoted by $\alpha$. The value of $\alpha$ is then used as a threshold for the later users, among which we will select the first one whose sensing value per unit time is no less than $\alpha$ and pay him/her $G$; other users all get empty schedules and zero payments (lines~\ref{ln:secralessg}-\ref{ln:secralessgend}). The schedules and payments assigned to the users are returned by vector $\vec{y}$ and vector $\vec{p}$, respectively.

  Clearly, \textbf{Algorithm}~\ref{alg:secra} provides a feasible solution to the MCS problem and satisfies IR. The truthfulness and competitive ratio of \textbf{Algorithm~\ref{alg:secra}} are given in \textit{\textbf{Theorem}~\ref{thm:secrtruthful}} and \textit{\textbf{Theorem}~\ref{thm:onlmchpartb}}, respectively:
  \begin{theorem}
    The online scheduling mechanism in \textbf{Algorithm~\ref{alg:secra}} is truthful.
    \label{thm:secrtruthful}
  \end{theorem}
  \begin{IEEEproof}
    As the users cannot control their arrival sequence, they also cannot control the value of $\alpha$. The first arrived $\lfloor \frac{n}{\mathbf{e}}\rfloor$ users are always assigned the empty schedule, so they always get the utility 0 no matter how they bid. Now consider any $A_i$ whose arrival order is greater than $\lfloor \frac{n}{\mathbf{e}}\rfloor$. If $A_i$ gets a non-empty schedule by bidding $\hat{d}_i$, then it is clear that he/she cannot benefit from lying, because his/her utility $G-\hat{d}_i\geq 0$ is the largest one he/she can possibly get. Otherwise if $A_i$ gets an empty schedule by bidding $\hat{d}_i$, then there are two cases we need to consider: (i) $G=0$ when $A_i$ arrives: In this case, $A_i$ will always get utility 0 no matter how he/she bids. (ii) $G>0$ when $A_i$ arrives: In this case we must have $\mu_i<\alpha$ or $\hat{d}_i>G$. If $\mu_i< \alpha$, then $A_i$ always gets utility 0 regardless of his/her bid. If $\mu_i\geq \alpha$ and $\hat{d}_i> G$, bidding any $d_i>G$ will always cause $A_i$ to get an empty schedule (hence the utility 0), while bidding $d_i\leq G$ would enable $A_i$ to get a non-empty schedule, but his/her utility $G-\hat{d}_i$ would be negative; hence he/she is better off bidding truthfully.
  \end{IEEEproof}

  \begin{theorem}
    If $\Lambda \geq \frac{R(\vec{y}^*)}{150}$ (where $\Lambda$ was defined in \textit{\textbf{Theorem}~\ref{thm:conditionalar}}), then \textbf{Algorithm~\ref{alg:secra}} has a $\mathcal{O}(1)$ competitive ratio with a constant probability.
    \label{thm:onlmchpartb}
  \end{theorem}
  Note that the competitive ratio stated in \textit{\textbf{Theorem}~\ref{thm:onlmchpartb}} is conditional. To rectify this problem, we propose a randomized mechanism shown in \textbf{Algorithm~\ref{alg:onlinemch}}, which runs \textbf{Algorithm~\ref{alg:secra}} with probability one half and runs lines~\ref{ln:ranstart}-\ref{ln:ranend} otherwise. Roughly speaking, the idea of lines~\ref{ln:ranstart}-\ref{ln:ranend} is the following: we assign the empty schedule to the first arrived $\xi$ users and use them as a random sample to guess the optimal solution $R(\vec{y}^*)$ (lines~\ref{ln:jlessxi}-\ref{ln:calransample}); then we use this guess to schedule the users coming afterwards (lines~\ref{ln:seteta}-\ref{ln:ranend}).
  %
%\begin{lemma}
%Let $X_1,X_2,...,X_k$ are independent random variables such that $\forall 1\leq i\leq k: \mathrm{Prob}\{X_i=1\}=p_i$ and $\mathrm{Prob}\{X_i=0\}=1-p_i$. Let $Y=\sum_{1\leq i\leq k}a_i X_i$ where $a_i : 1\leq i\leq k$ are non-negative real numbers. Let $m$ be any positive number which satisfies $m\geq \max\{a_i | 1\leq i \leq k\}$. Suppose $\mathbb{E}(Y)\leq S$, then for any $\delta\geq 0$ we have:
%\begin{eqnarray}
%\mathrm{Prob}\{Y\geq (1+\delta)S\} \leq \left( \frac{\mathbf{e}^\delta}{(1+\delta)^{1+\delta}} \right) ^{{\frac{S}{m}}} \nonumber
%\end{eqnarray}
%\label{lma:chernoff}
%\end{lemma}
%
%\begin{lemma} [Chernoff Bound]
%Suppose that $X_1,X_2,...,X_j$ are independent random variables which satisfy $ X_i\in [0,\theta]$ $(\forall 1\leq i\leq j)$. Let $H=\sum_{i=1}^k X_i$. Then for any $\delta>0$, we have:
%\begin{eqnarray}
%\mathrm{Prob}\{H\geq (1+\delta)\mathbb{E}(H)\} \leq \left( \frac{\mathbf{e}^\delta}{(1+\delta)^{1+\delta}} \right) ^{{\frac{\mathbb{E}(H)}{\theta}}} \nonumber
%\end{eqnarray}
%and
%\begin{eqnarray}
%\mathrm{Prob}\{H\leq (1-\delta)\mathbb{E}(H)\} \leq e ^{-{\frac{\delta^2\mathbb{E}(H)}{2\theta}}} \nonumber
%\end{eqnarray}
%\end{lemma}
  %
  \begin{algorithm}[htb]
%   \KwIn{$G, \vec{b}, \mathcal{A}, T$}
%   %
%   \KwOut{$S, w$}
    %
%   \lFor{$i\leftarrow 1$ \KwTo $n$}{
%       $z_i\leftarrow \emptyset,p_i\leftarrow 0$
%   }
    Generate a random number $o$ from the uniform distribution on the interval [0, 1];\\
    \If{$o\leq \frac{1}{2}$}{
        Let $\xi$ be a random number generated from the binomial distribution $B(n,1/2)$ \label{ln:ranstart} \\
%       Get the first $\xi$ input users and reject them;\\
        $M\leftarrow G$;~~~$j\leftarrow 0$;~~~$\mathcal{A}_\xi \leftarrow \emptyset$ \\
        \SetKwBlock{Upon}{upon}{end}
        \Upon($A_i$'s arrival){
%       \While {a user with id $i$ comes} {
            $j\leftarrow j+1$\\
            \eIf {$j\leq \xi$}{ \label{ln:jlessxi}
                $y_i\leftarrow \emptyset$;~~~$p_i\leftarrow 0$;~~~$\mathcal{A}_\xi \leftarrow \mathcal{A}_\xi \bigcup \{A_i\}$ \\
                \If {$j=\xi$}{
                    $\vec{r}(\vec{b}_\xi) \leftarrow \mathsf{ApproxMCS}(\xi, G, \vec{b}_\xi, \mathcal{A}_\xi)$ \label{ln:calransample} \\
%                   Run \textbf{Algorithm~\ref{alg:assc}} to calculate the sensing schedule vector $\vec{r}$ for the $\xi$ users that have arrived\\
                }
            }{
                $\eta\leftarrow 5G\cdot\mu_i /R(\vec{r})$;~~~$F_i\leftarrow T_i-\bigcup_{\ell:\kappa_i=\kappa_\ell}y_\ell$\\ \label{ln:seteta}
                %$\eta_1\leftarrow 5G\cdot\mu_i /R(\vec{r})$\\
                %$\eta_2\leftarrow b_i\cdot \lfloor \frac{M}{b_i}\rfloor + \int_{b_i}^{\eta_1} \lfloor \frac{M}{v}\rfloor \mathrm{d}v$\\
                \eIf{$d_i\leq \eta \bigwedge \eta\cdot |F_i|\leq M \bigwedge |F_i|\geq 1$}{ \label{ln:allocationjudge}
                %\eIf{$b_i\leq \eta_1\wedge \eta_2\leq M\wedge |Z|\geq 1$}{
                    $y_i\leftarrow F_i$;~~~$p_i\leftarrow \eta\cdot |F_i|$;~~~$M\leftarrow M-p_i$ \label{ln:payend}\\
                    %$z_i\leftarrow$ any sub-period of $Z$ with $\lfloor M/b_i\rfloor$ time slots\\
                    %$p_i\leftarrow \eta_2,M\leftarrow M-p_i$\\
                }{
                    $y_i\leftarrow \emptyset$;~~~$p_i\leftarrow 0$ \label{ln:ranend}
                }
            }
        }
%        \For{$i\leftarrow \xi+1$ \KwTo $n$}{
%
%        }
    }
    \lElse{
         Run \textbf{Algorithm~\ref{alg:secra}} to get $(\vec{y},\vec{p})$\\
%        $j\leftarrow \arg\max_{1\leq i\leq n}\left(\mu_i\cdot \min\{\lfloor\frac{G}{b_i}\rfloor,|T_i|\}\right)$\\
%        Find any sub-period in $T_j$ with $\min\{|T_j|,\lfloor\frac{G}{b_j}\rfloor\}$ time slots as $A_j$'s sensing schedule\\
%        Pay $A_j$ the amount $G$ and pay others zero.\\
    }
    \Return{$(\vec{y},\vec{p})$}
    \caption{A Randomized Online Mechanism} \label{alg:onlinemch}
  \end{algorithm}
  It can be seen from lines~\ref{ln:allocationjudge}-\ref{ln:payend} that \textbf{Algorithm~\ref{alg:onlinemch}} satisfies IR and provides a feasible solution to the MCS problem. The truthfulness of \textbf{Algorithm~\ref{alg:onlinemch}} is proven in \textit{\textbf{Theorem}~\ref{thm:semitruthful}}:
  \begin{theorem}
    The online scheduling mechanism shown in \textbf{Algorithm}~\ref{alg:onlinemch} is truthful.
    \label{thm:semitruthful}
  \end{theorem}
  \begin{IEEEproof}
    We have proved in \textit{\textbf{Theorem}~\ref{thm:secrtruthful}} that \textbf{Algorithm~\ref{alg:secra}} is truthful, so we only need to prove that lines~\ref{ln:ranstart}-\ref{ln:ranend} are truthful given that users strategically report their sensing costs. Note that the users cannot control their arrival order as well as the value of $R(\vec{r})$, hence the first arrived $\xi$ users always get utility $0$ no matter how they bid. Now consider any user $A_i$ who arrives afterwards. If $A_i$ gets a non-empty schedule (i.e., $|y_i|\neq 0$) by bidding $\hat{d}_i$, then we know that $\hat{d}_i\leq \eta$ and $A_i$ gets the utility $(\eta-\hat{d}_i)\cdot |F_i|\geq 0$, which remains the same if $A_i$ bids any $d_i\leq \eta$. If $A_i$ bids $d_i> \eta$, then his/her utility will be $0$, so he/she is better off bidding his/her true value. Otherwise if $|y_i|= 0$ when $A_i$ bids $\hat{d}_i$, then at least one of the following conditions holds: (i) $\eta\cdot |F_i|> M$; (ii) $|F_i|=0$; (iii) $\hat{d}_i>\eta$. If (i) or (ii) holds, $A_i$'s utility remains $0$ no matter how he/she bids. If (iii) holds, then bidding $d_i\leq \eta$ may get $A_i$ assigned a non-empty schedule, but in that case the utility of $A_i$ would be $(\eta-\hat{d}_i)\cdot |F_i|<0$, hence he/she is still better off bidding the true value $\hat{d}_i$.
  \end{IEEEproof}

  Finally, the competitive ratio and time complexity of \textbf{Algorithm~\ref{alg:onlinemch}} are given in \textit{\textbf{Lemma}~\ref{lma:deltabound}-\ref{lma:onlmchparta}} and \textit{\textbf{Theorem}~\ref{thm:artimcom}}:

%Let $\Lambda=\max_{1\leq i\leq n}\{\mu_i\cdot \min(|T_i|,\lfloor \frac{G}{b_i}\rfloor)\}$. Let $\vec{w}^*=(w^*_1,...,w^*_n)$ be an optimal solution. Let $(\sigma_1,\sigma_2,...,\sigma_n)$ be the actual arrival sequence of the users's ids, which is a permutation of $\{1,2,...,n\}$. Let $\Delta_1=\sum_{i=1}^\xi \mu_{\sigma_i}\cdot |w^*_{\sigma_i}|$ and $\Delta_2=\sum_{i=\xi+1}^n \mu_{\sigma_i}\cdot |w^*_{\sigma_i}|$. Then we have:

%\begin{lemma}
%Any user is among the first $\xi$ arrived users with probability of $\frac{1}{2}$ independently of other users.
%\end{lemma}
%\begin{proof}
%For any $1\leq i\leq n$, we have
%\begin{eqnarray}
%&&\mathrm{Prob}\{i\in\{\sigma_1,\sigma_2,...,\sigma_{\xi}\}\} \nonumber\\
%&=& \sum_{j=1}^n \mathrm{Prob}\{\sigma_j=i\}\cdot \mathrm{Prob}\{\xi\geq j\}\\
%&=& \frac{1}{n}\mathbb{E}(\xi)=\frac{1}{2} \nonumber
%\end{eqnarray}
%\end{proof}

  \begin{lemma}
    Let $(\sigma_1,\sigma_2,...,\sigma_n)$ be the actual arrival sequence of the users' indices, which is a permutation of $\{1,2,...,n\}$. Let $\Delta_1=\sum_{i=1}^\xi \mu_{\sigma_i}\cdot |y^*_{\sigma_i}|$ and $\Delta_2=\sum_{i=\xi+1}^n \mu_{\sigma_i}\cdot |y^*_{\sigma_i}|$. If $\Lambda \leq \frac{R(\vec{y}^*)}{150}$, then $\Delta_1\geq {R(\vec{y}^*)}/{3}$ and $\Delta_2\geq {R(\vec{y}^*)}/{4}$ hold at the same time with constant probability.
    %\begin{eqnarray}
    %\mathrm{Prob}\{\Delta_1\geq \frac{R(\vec{w}^*)}{3}\wedge \Delta_2\geq \frac{R(\vec{w}^*)}{4}\}\geq 0.983 \nonumber
    %\end{eqnarray}
    \label{lma:deltabound}
  \end{lemma}
  \begin{lemma}
    If $\Lambda \leq \frac{R(\vec{y}^*)}{150}$, then $\frac{\Delta_1}{5}\leq R(\vec{r})\leq R(\vec{y}^*)$.
    \label{lma:boundapx}
  \end{lemma}
  %
%\begin{lemma}
%If $\Lambda \leq \frac{R(\vec{w}^*)}{150}$, then $\frac{R(\vec{w}^*)}{15}\leq R(\vec{r})\leq 4\Delta_2$ with a probability of at least $0.983$
%\end{lemma}
%\begin{proof}
%Let $opt_1$ be the optimal value of sensing scheduling for the first half of the input $\{A_{\sigma_1},...,A_{\sigma_{\xi}}\}$. Using Lemma~\ref{lma:conditionalar} with $\epsilon=\frac{e-1}{2e}-\frac{1}{150}$, we get
%\begin{eqnarray}
%R(\vec{r})\geq \frac{2}{3}\cdot \left(\frac{e-1}{2e}-\frac{1}{150}\right)\cdot opt_1 \geq 0.2\cdot opt_1 \nonumber
%\end{eqnarray}
%As $opt_1\geq \Delta_1$ and $\Delta_1\geq R(\vec{w}^*)/3$, we get $R(\vec{r})\geq R(\vec{w}^*)/15$. On the other hand,
%\begin{eqnarray}
%R(\vec{r})\leq  opt_1\leq  R(\vec{w}^*)
%\end{eqnarray}
%As $\Delta_2\geq \frac{R(\vec{w}^*)}{4}$, we know $R(\vec{r})\leq 4\Delta_2$.
%%Let $(\sigma_1,\sigma_2,...,\sigma_n)$ be a permutation of $(1,2,...,n)$ such that $|\rho_{\sigma_1}|\cdot \mu_{\sigma_1}\geq |\rho_{\sigma_2}|\cdot \mu_{\sigma_2}\geq...\geq |\rho_{\sigma_n}|\cdot \mu_{\sigma_n}$. Let $\{Y_1,Y_2,...,Y_n\}$ be a set of random variables such that $Y_{\sigma_i}=|\rho_{\sigma_i}|\cdot \mu_{\sigma_i}$ if $1\leq \sigma_i\leq \xi$, and $Y_{\sigma_i}=0$ otherwise. Let $Y=\sum_{i=1}^k Y_{i}$ and $\bar{Y}=R(\vec{w}^*)-Y$. So we have $Y=\sum_{i=1}^\xi |\rho_{i}|\cdot \mu_i$ and $\bar{Y}=\sum_{i=\xi+1}^n |\rho_{i}|\cdot \mu_i$
%\end{proof}
  %
  \begin{lemma}
    If $\Lambda \leq \frac{R(\vec{y}^*)}{150}$, then the solution output by lines~\ref{ln:ranstart}-\ref{ln:ranend} of \textbf{Algorithm~\ref{alg:onlinemch}} has an $\mathcal{O}(1)$ competitive ratio with constant probability.
  \label{lma:onlmchparta}
  \end{lemma}
  \begin{theorem}
    The competitive ratio and worst-case time complexity of \textbf{Algorithm~\ref{alg:onlinemch}} are $\mathcal{O}(1)$ and $\mathcal{O}(n^2)$, respectively.
    \label{thm:artimcom}
  \end{theorem}

  \begin{figure*}[htb]
    %\vspace{-.5ex}
    \begin{center}
        \subfigure[Revenue changing with user number.]{\label{fig:RevenueVsUser}\includegraphics[width=.325\textwidth]{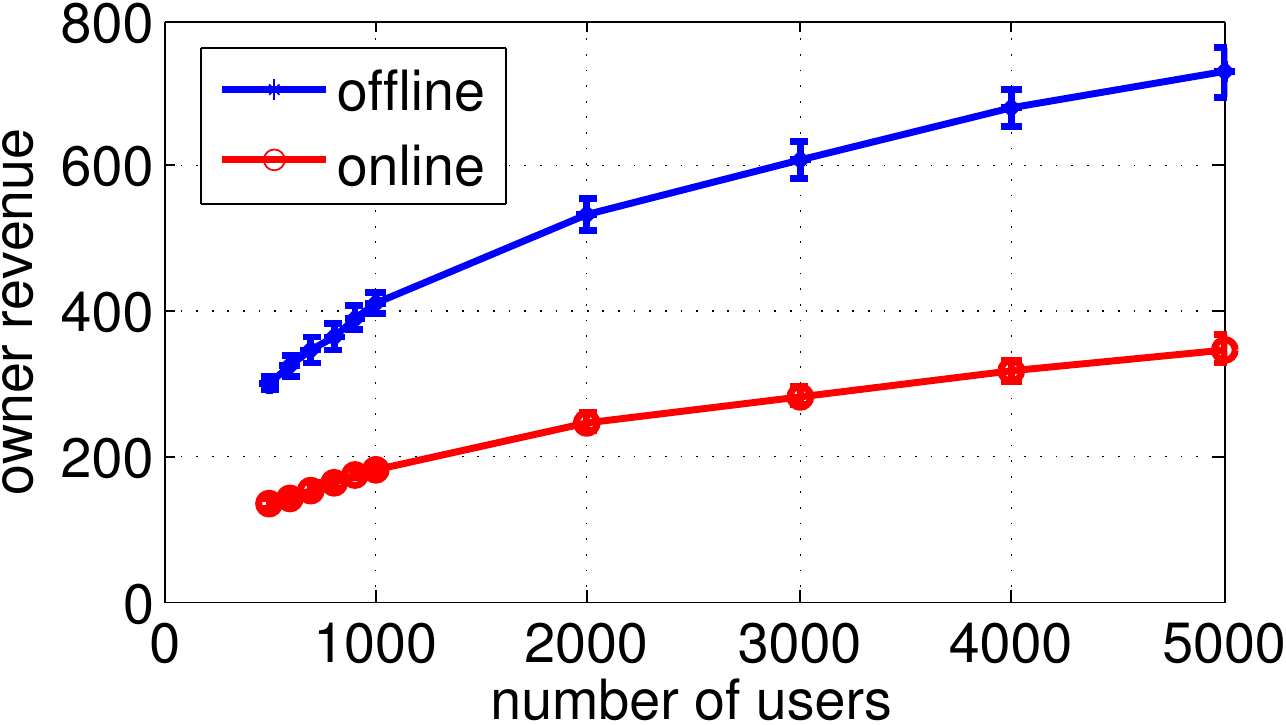}}
        \subfigure[Revenue changing with budget.]{\label{fig:RevenueVsBudget}\includegraphics[width=.32\textwidth]{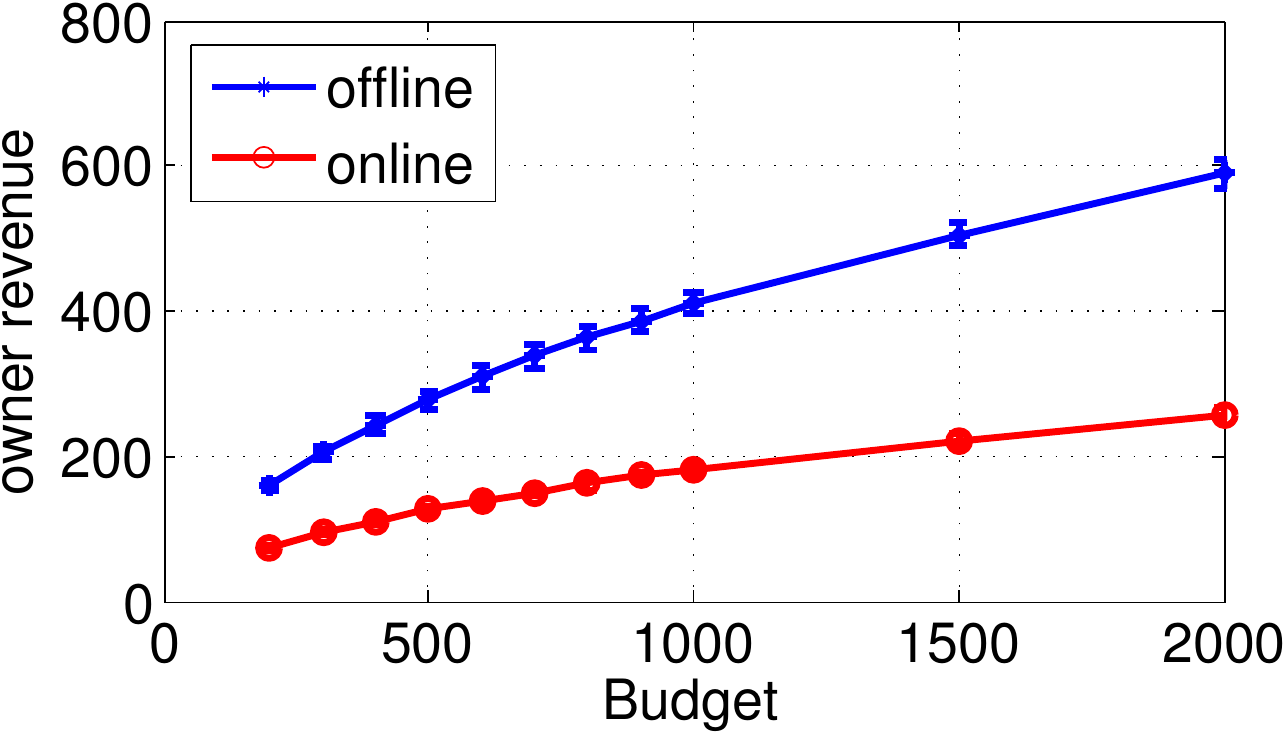}}
        \subfigure[Revenue changing with task number]{\label{fig:RevenueVsTask}\includegraphics[width=.32\textwidth]{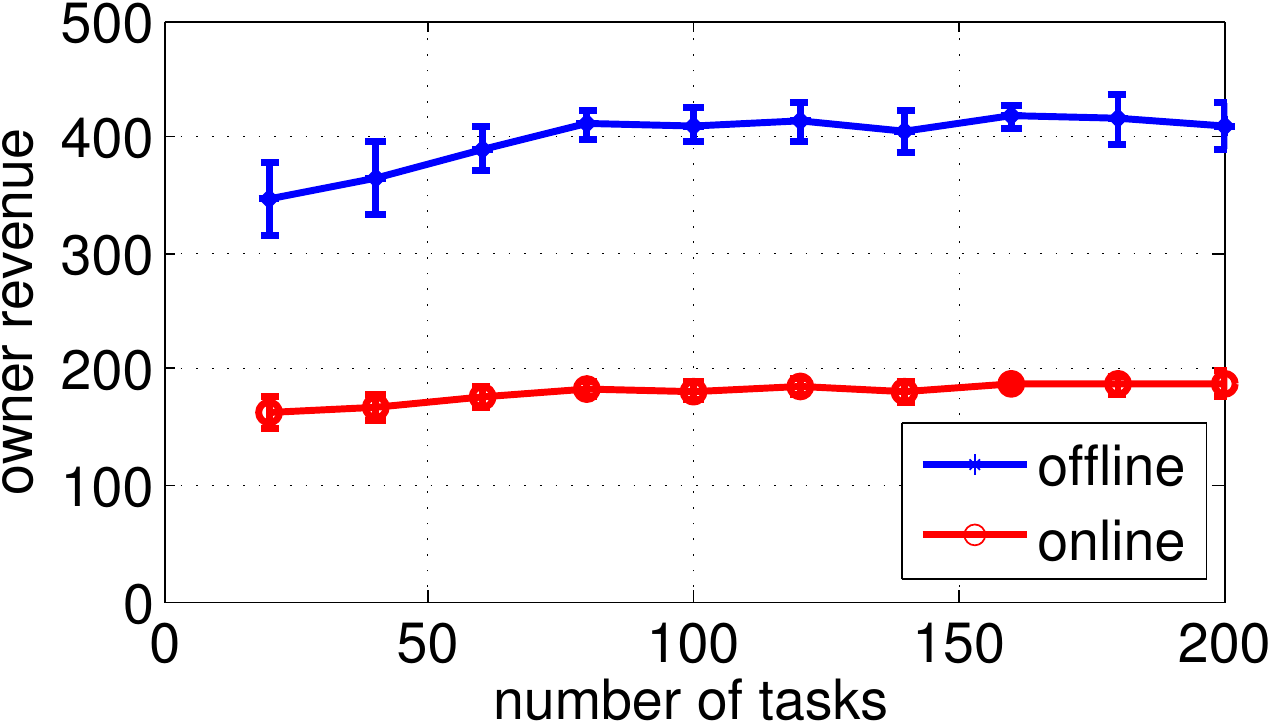}}
    \end{center} \vspace{-1.5ex}
    \caption{The owner's revenue as functions of (a) the number of users, (b) the total budget, and (c) the number of tasks.}
    \label{fig:revenue} %\vspace{-1ex}
  \end{figure*}
  \begin{figure*}[htb]
    %\vspace{-.5ex}
    \begin{center}
        \subfigure[Payment changing with budget.]{\label{fig:payment}\includegraphics[width=.325\textwidth,height=.195\textwidth]{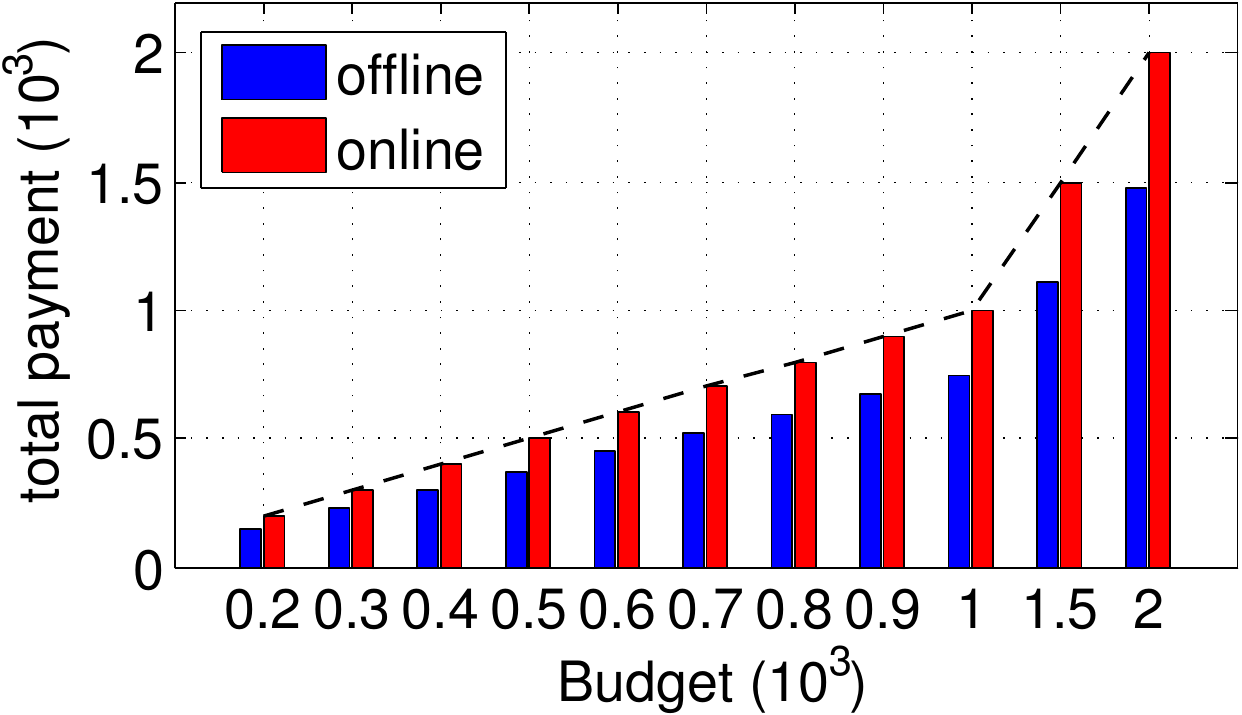}}
        \subfigure[IR validation for \textsf{offline}.]{\label{fig:IRoffline}\includegraphics[width=.312\textwidth]{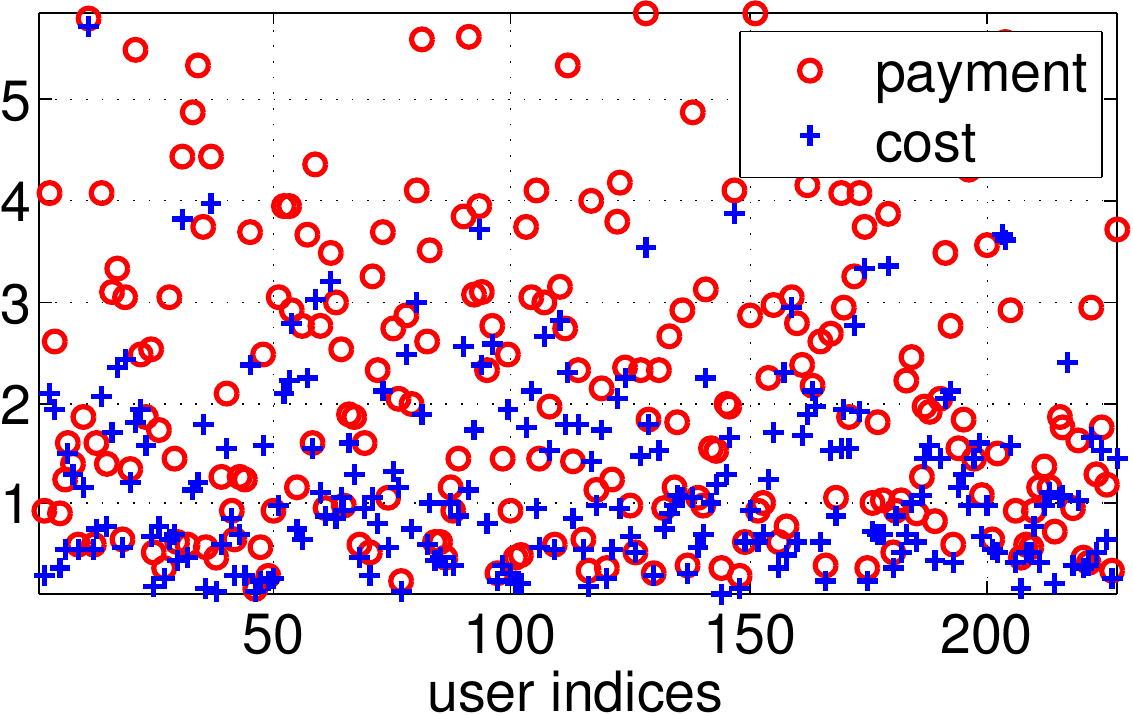}}
        \subfigure[IR validation for \textsf{online}.]{\label{fig:IRonline}\includegraphics[width=.32\textwidth]{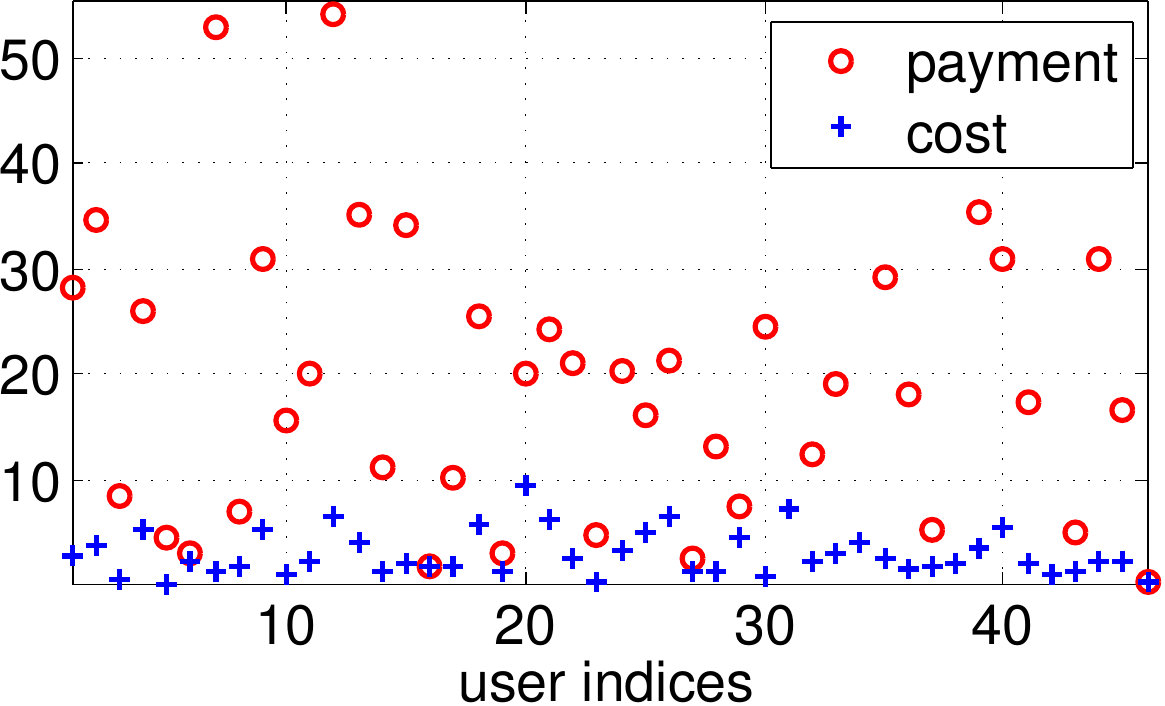}}
    \end{center} \vspace{-1.5ex}
    \caption{Feasibility check in terms of total payment and IR.}
    \label{fig:IR} \vspace{-2ex}
  \end{figure*}

\section{Simulations} \label{sec:simu}
  We conduct extensive simulations to evaluate the performance of our truthful scheduling mechanisms. The objective of our simulations is to corroborate the correctness and effectiveness of our mechanisms in terms of various metrics (including owner revenue, total payment, truthfulness and IR) under different parameter settings (such as the number of users or tasks, as well as the budget). Since we are the first, to the best of our knowledge, to study the MCS problem (see the discussions in Sec.~\ref{sec:intro} and the definition in Sec.~\ref{sec:model}), we can only make comparisons between our own algorithms in the simulations. For brevity, we hereby denote our two main algorithms \textbf{Algorithm}~\ref{alg:mch} and \textbf{Algorithm}~\ref{alg:onlinemch} by ``\textsf{offline}'' and ``\textsf{online}'', respectively.

  \subsection{Default Settings} \label{sec:simgeneral}
    We randomly generate the sensing values of tasks, the number of users and the users' private values. More specifically, the sensing cost per unit time (i.e., $\hat{d}_i$) of any user $A_i$ is generated randomly from the uniform distribution $U[0.1,1.1]$, and the sensing value per unit time $u_i$ of any task $K_i$ is also generated from the same distribution (i.e., both are bounded away from 0). The earliest available time point $\hat{s}_i$ of any user $A_i$ is generated randomly from $U[0,100]$, whereas the length of $A_i$'s available time period is generated randomly from $U[0,10]$. Both the number of users $n$ and the budget $G$ are set to 1000; the number of tasks $m$ is set 100; and each user selects only one task, with equal probability out of all tasks. All our simulations follow these default settings unless otherwise stated.

  \subsection{Owner Revenue}
    We study the owner revenue achieved by our mechanisms under different user number, budget and task number in Fig.~\ref{fig:revenue}. For each data point, we perform 100 simulations with random inputs and we plot the average value and standard deviation. In general, \textsf{offline} always works better than \textsf{online}. This is natural because \textsf{online} faces a harsher condition that a schedule has to be determined for a user upon his/her arrival. In particular, with the information on all users, \textsf{offline} can leverage a sorting based on $\prec$ to optimize the performance, whereas \textsf{online} do not have this privilege.

    In Fig.~\ref{fig:RevenueVsUser}, we study the impact of the number of users on the owner revenue, by scaling the number of users from 500 to 5000 with an increment of 100 (below 1000) and of 1000 (beyond 1000). The owner revenues of both \textsf{offline} and \textsf{online} increase with the number of users. This can be explained by the reason that, as the diversities of both the users' sensing costs and available time periods increase with the number of users, the degree of freedom in finding schedules is enlarged, which in turn results in larger revenues. The same trend is also shown in Fig.~\ref{fig:RevenueVsBudget}, where we fix the number of users to 1000 but scale the budget $G$ from 200 to 2000 with an increment of 100 (before 1000) and of 500 (after 1000). This is rather straightforward to understood because a higher budget allows the algorithms to schedule users whose sensing costs are higher and hence cannot be afforded under a lower budget.

    In Fig.~\ref{fig:RevenueVsTask}, we fix the user number to 1000 and increase the number of tasks from 20 to 200 with an increment of 20. It can be seen that the owner revenues obtained by both our mechanisms slightly increase when the number of tasks increases. This can be explained by the reason that, when the number of tasks increases, the number of users that can perform each task tend to decrease, which results in less overlapping available time periods and higher revenue. %due to the scheduling methods in \textbf{Algorithm}~\ref{alg:mch} and \textbf{Algorithm}~\ref{alg:onlinemch}.
    Obviously, this effect is less direct than that from either increasing user number or budget, so the resulting improvement to the revenue is also marginal.

  \subsection{Solution Feasibility and Individual Rationality}
    We verify the feasibility and IR of the solutions output by our algorithms in this section. We first show that the total payment is always no more than the budget, then we use two examples to demonstrate that IR is also guaranteed, i.e., each user gets a payment higher than his/her cost. In Fig.~\ref{fig:payment}, we scale the budget in the same way as Fig.~\ref{fig:RevenueVsBudget}, and we show the maximum total payment for each case. Apparently, the budget has never been surpassed. Again, \textsf{offline} is shown to be superior to \textsf{online}: it results in lower total payments.

    In Fig.~\ref{fig:IRoffline} and~\ref{fig:IRonline}, we demonstrate IR using the outputs from \textsf{offline} and \textsf{online}, respectively. We plot the sensing costs and payments only for users with non-zero payments. IR of our mechanisms can be immediately seen: a payment is always greater than the corresponding cost. We can also see that more users are assigned non-empty schedules by \textsf{offline}, which, to some extent, explains the observation made for Fig.~\ref{fig:revenue} that \textsf{offline} always achieves a higher revenue than \textsf{online}.

  \subsection{Truthfulness}
    We verify the truthfulness of both \textsf{offline} and \textsf{online} by arbitrarily picking up a few users and checking their utilities under different bidding values.

    %A thorough verification has to go through all users and all possible bids, but we can only demonstrate a few cases due to limited space.

\begin{figure}[htb]
        %\vspace{-.5ex}
        \begin{center}
        \subfigure[Truthfulness on the sensing cost.]{\label{fig:varydi}\includegraphics[width=.24\textwidth]{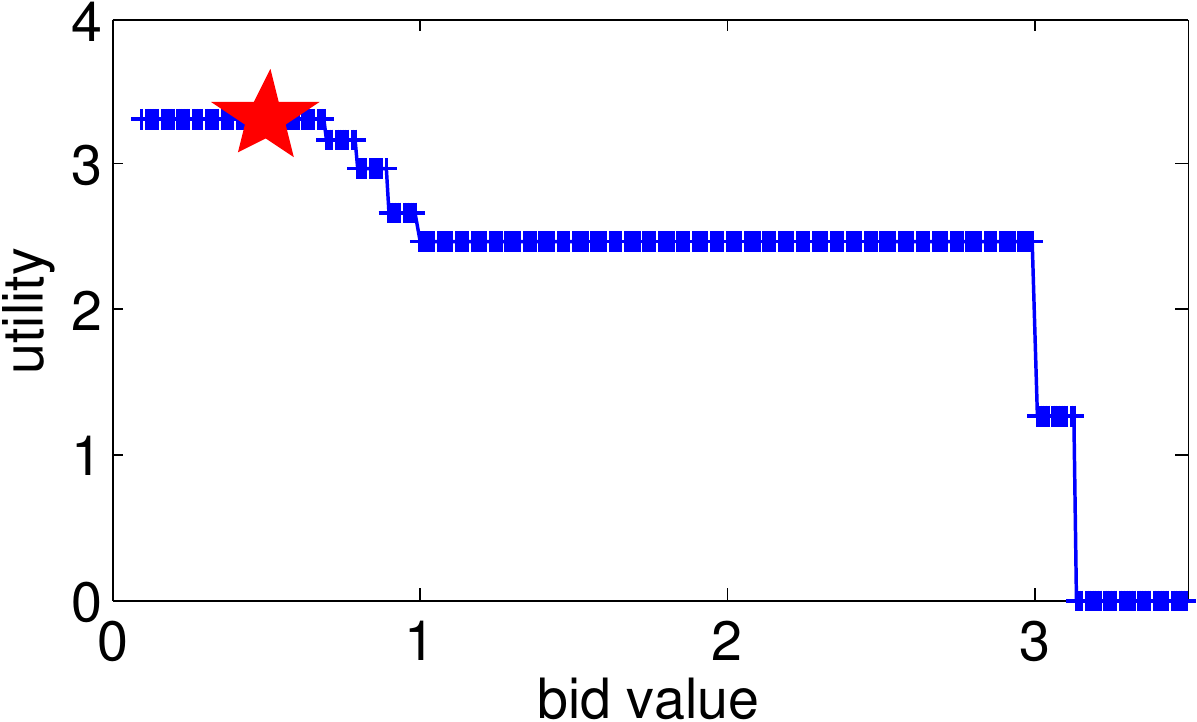}}
%        \subfigure[offline truthfulness validation on bid time period.]{\label{fig:gmimap}\includegraphics[width=.32\textwidth]{figures/offline_truth_vali_time}}
        \subfigure[Truthfulness on the available time.]{\label{fig:varyTi}\includegraphics[width=.24\textwidth]{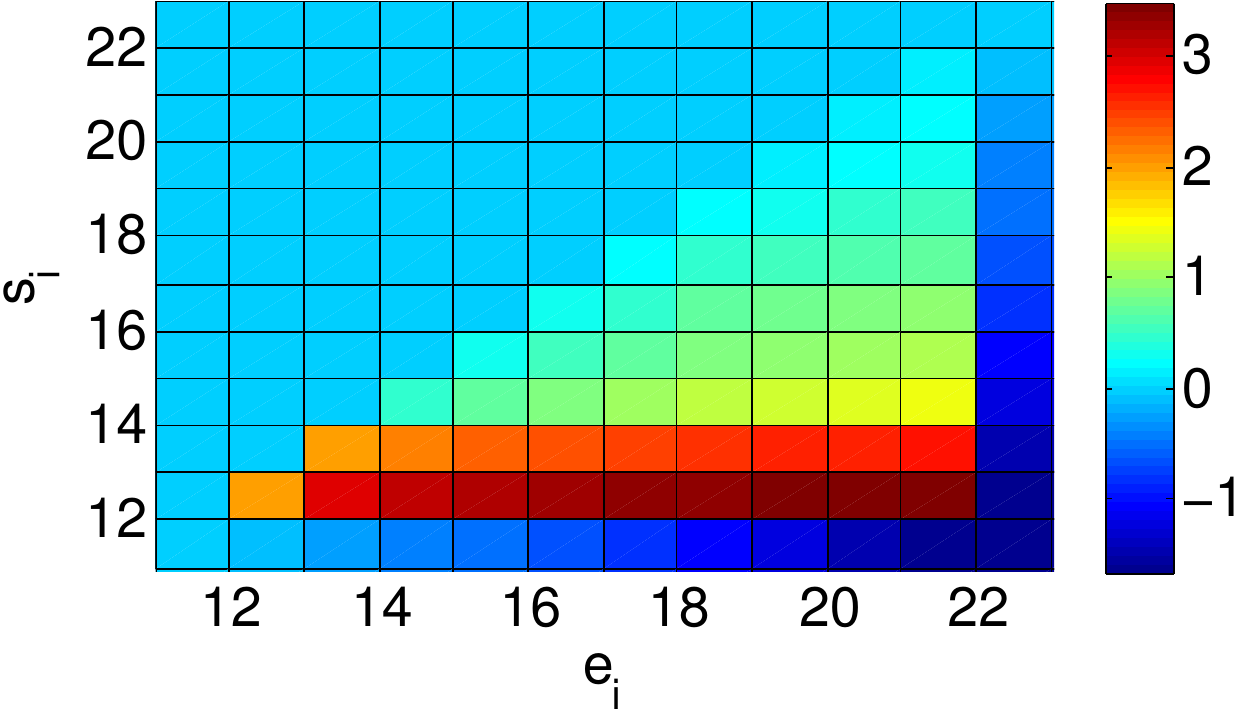}}
        \end{center} \vspace{-1ex}
        \caption{Truthfulness validation for \textsf{offline}.}
      \label{fig:offlinetuthfulness} \vspace{-2ex}
      \end{figure}

    We first study the truthfulness of \textsf{offline} by Fig.~\ref{fig:offlinetuthfulness}. We arbitrarily pick a user $A_i$ whose true values are $\hat{d}_i=0.5$, $\hat{s}_i=12$ and $\hat{e}_i=22$, then we change $A_i$'s bid (with other users' bids fixed) to see how $A_i$'s utility changes. Since we cannot draw a 4-dimensional chart here, we show the results by two figures. In Fig.~\ref{fig:varydi}, the bid of $A_i$'s available time period is fixed to $[12,22]$, and we scale $A_i$'s bid on his sensing cost from 0.1 to 3.3 with an increment of 0.1. Indeed, bidding the true sensing cost (shown by the red pentagram) allows $A_i$ to maximize his/her utility. In Fig.~\ref{fig:varyTi}, we fix $A_i$'s bid on his sensing cost to 0.5, but varies $A_i$'s bid on his earliest and latest available time points. Again, $A_i$'s utility is maximized when he/she bids his/her true value $[12,22]$. These demonstrate that $A_i$ has no incentive to deviate from bidding his/her true values.

\begin{figure}[htb]
        %\vspace{-.5ex}
        \begin{center}
        \subfigure[Truthfulness for a user $A_i$ who gets an empty schedule.]{\label{fig:loser}\includegraphics[width=.24\textwidth]{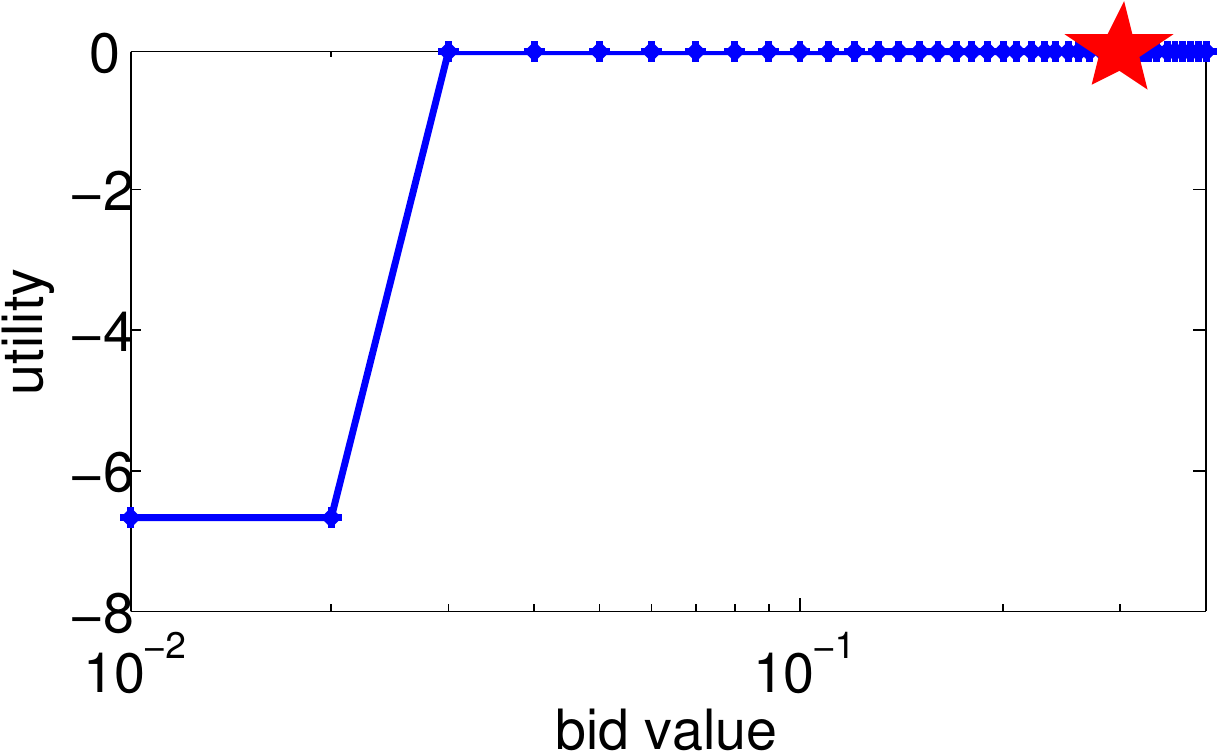}}
        \subfigure[Truthfulness for a user $A_i$ who gets a non-empty schedule.]{\label{fig:winner}\includegraphics[width=.24\textwidth]{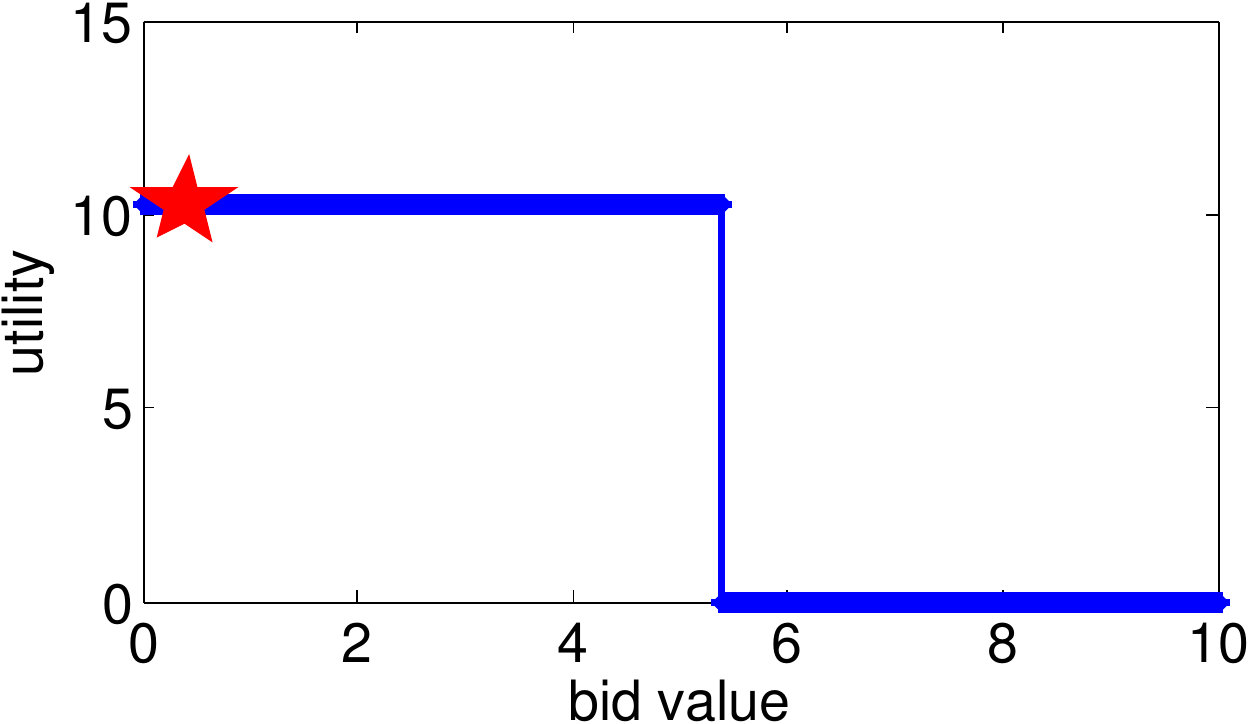}}
        \end{center} \vspace{-1ex}
        \caption{Truthfulness validation for \textsf{online}.}
      \label{fig:onlinetruthfulness} \vspace{-2ex}
      \end{figure}

    Similarly, we study the truthfulness of \textsf{online} by Fig.~\ref{fig:onlinetruthfulness}. In Fig.~\ref{fig:loser}, we pick an arbitrary user whose true sensing cost is 0.3 and who is assigned an empty schedule by \textsf{online}. Then we scale this user's bid on his sensing cost from 0.01 to 0.4 with an increment of 0.01. The user indeed achieves his/her maximum utility 0 by bidding his true sensing cost 0.3. In Fig.~\ref{fig:winner}, we pick another user (in another simulation) whose true sensing cost is 0.4 and who is assigned a non-empty schedule by \textsf{online}. Then we scale this user's bid from 0.01 to 10 with an increment of 0.01. Again, the result shows that the user's utility is maximized when he/she bids his/her true sensing cost.

    % Although we only choose a few users to demonstrate that bidding their true values are their dominant strategy, the same situation applies to every other users. Therefore, both \textsf{offline} and \textsf{online} are truthful mechanisms.

\section{Related Work} \label{sec:rw}
%
% To the best of our knowledge, there are currently very few proposals on incentive mechanism design for mobile sensing.
% and the most close work to ours is \cite{Yang2012}.
  Mobile crowdsensing involves using (human carried) smartphones to gather data in a much larger scale than what can be done in conventional ways, either through autonomous phone sensing or by further demanding active human participation\cite{GantiYL2011,Khan2013}. While the developments on mobile crowdsensing applications are plentiful, only a few proposals have started on studying how to incentivize participation to such applications until very recently~\cite{Lee2010,Jaimes2012,Duan2012,Yang2012,Qinghua2013}.

  Duan \textit{et al}.\ \cite{Duan2012} have proposed incentive mechanisms to motivate collaboration in mobile crowdsensing based on Stackelberg games and contract theory. However, the mechanisms provided in~\cite{Duan2012} require either the complete information or the prior distributions of users' private types, hence are not prior-free mechanisms as those in our work. Yang \textit{et al}.\ \cite{Yang2012} suggest both a platform-centric model and a user-centric model for mobile sensing. They also use a Stackelberg game to design incentive mechanisms for the platform-centric model, and use auction theory to design truthful mechanisms for the user-centric model; nevertheless the truthful auction mechanisms provided in~\cite{Yang2012} are only for single-parameter users and only run in an offline manner. Moreover, no theoretical performance ratios are provided for them in~\cite{Yang2012}.
  % Koutsopoulos~\cite{Koutsopoulos2013} has proposed a single-parameter offline mechanism to minimize the compensation costs in mobile crowdsensing while maintaining a satisfying service level to subscribers. However, the \textcolor{red}{Bayesian} mechanisms proposed in~\cite{Koutsopoulos2013} require the \textcolor{red}{prior distributions} of users' private types (which is hard to obtain in practice), hence are not prior-free mechanisms as those in our work.
  Some other issues such as pricing, coverage, and privacy of mobile crowdsensing have also been studied by the work in~\cite{Lee2010,Jaimes2012,Qinghua2013}, but these proposals are either not based on a game theoretical perspective or have not considered important game-theoretic issues such as truthfulness and IR.
  % have studied Another work~\cite{Qinghua2013} has proposed privacy-aware incentive schemes for mobile sensing, but not from a game-theoretic perspective.
  Most importantly, none of the work in~\cite{Lee2010,Jaimes2012,Duan2012,Yang2012,Qinghua2013} has considered the special time scheduling problem arising from the mobile crowdsensing paradigm, hence their problem definitions are totally different from ours.
%
% Although our sensing model has some similarities with the user-centric model in~\cite{Yang2012}, the problem we have studied in this paper is totally different from the problem studied in~\cite{Yang2012}, because  the sensing scheduling problem was not considered in~\cite{Yang2012} under the user-centric model. We also note that only offline mechanisms are provided in~\cite{Yang2012} and no theoretical bounds are given there.

  There also exist proposals on designing approximation algorithms or truthful mechanisms for job-scheduling on parallel machines, such as~\cite{Lenstra1990,Christodoulou2010,Dhangwatnotai2011,Koutsoupias2013}. However, these proposals focus on the problem of minimizing the scheduling makespan, which is a totally different goal from ours. Besides, all the mechanisms in this line could entail an arbitrarily large payment to ensure truthfulness. Finally, the frugal or budget-feasible mechanism design problems have been studied in~\cite{Karlin2005,Singer2010,Chen2011,Badanidiyuru2012}, but these proposals only aim at designing single-parameter mechanisms for allocating indivisible goods, which is very different from the scheduling problems studied in this paper.

\section{Conclusion} \label{sec:con}
  We have studied incentive mechanisms for a novel scheduling problem (the MCS problem) arising from the mobile crowdsensing paradigm, where an application owner pays the sensor carriers and schedules their sensing time based on their bids to maximize the total sensing value. We have proved the NP-hardness of the MCS problem, and proposed polynomial-time approximation mechanisms for it that run both offline and online. We also have proved that our mechanisms have $\mathcal{O}(1)$ performance ratios and satisfies game-theoretic properties including individual rationality and truthfulness. The effectiveness of our approach has been corroborated by the simulation results. To the best of our knowledge, we are the first to study the mechanism design problems for the mobile crowdsensing scheduling problem.

\bibliographystyle{IEEEtran}

\bibliography{IEEEabrv,MyLib}

\begin{thebibliography}{10}
\providecommand{\url}[1]{#1}
\csname url@samestyle\endcsname
\providecommand{\newblock}{\relax}
\providecommand{\bibinfo}[2]{#2}
\providecommand{\BIBentrySTDinterwordspacing}{\spaceskip=0pt\relax}
\providecommand{\BIBentryALTinterwordstretchfactor}{4}
\providecommand{\BIBentryALTinterwordspacing}{\spaceskip=\fontdimen2\font plus
\BIBentryALTinterwordstretchfactor\fontdimen3\font minus
  \fontdimen4\font\relax}
\providecommand{\BIBforeignlanguage}[2]{{%
\expandafter\ifx\csname l@#1\endcsname\relax
\typeout{** WARNING: IEEEtran.bst: No hyphenation pattern has been}%
\typeout{** loaded for the language `#1'. Using the pattern for}%
\typeout{** the default language instead.}%
\else
\language=\csname l@#1\endcsname
\fi
#2}}
\providecommand{\BIBdecl}{\relax}
\BIBdecl

\bibitem{GantiYL2011}
R.~Ganti, F.~Ye, and H.~Lei, ``{Mobile Crowdsensing: Current State and Future
  Challenges},'' \emph{IEEE Communications Magazine}, vol.~6, no.~11, pp.
  32--39, 2011.

\bibitem{Khan2013}
W.~Khan, Y.~Xiang, M.~Aalsalem, and Q.~Arshad, ``{Mobile Phone Sensing Systems:
  A Survey},'' \emph{IEEE Communications Surveys \& Tutorials}, vol.~15, no.~1,
  pp. 402 --427, 2013.

\bibitem{Lee2010}
J.-S. Lee and B.~Hoh, ``{Dynamic Pricing Incentive for Participatory
  Sensing},'' \emph{Elsevier Pervasive and Mobile Computing}, vol.~6, no.~6,
  pp. 693--708, 2010.

\bibitem{Jaimes2012}
L.~Jaimes, I.~Vergara-Laurens, and M.~Labrador, ``{A Location-Based Incentive
  Mechanism for Participatory Sensing Systems with Budget Constraints},'' in
  \emph{Proc. of the 10th IEEE PerCom}, 2012, pp. 103--108.

\bibitem{Duan2012}
L.~Duan, T.~Kubo, K.~Sugiyama, J.~Huang, T.~Hasegawa, and J.~Walrand,
  ``{Incentive mechanisms for smartphone collaboration in data acquisition and
  distributed computing},'' in \emph{Proc. of the 31th IEEE INFOCOM}, 2012, pp.
  1701--1709.

\bibitem{Yang2012}
D.~Yang, G.~Xue, X.~Fang, and J.~Tang, ``{Crowdsourcing to Smartphones:
  Incentive Mechanism Design for Mobile Phone Sensing},'' in \emph{Proc. of the
  18th ACM MobiCom}, 2012, pp. 173--184.

\bibitem{Qinghua2013}
Q.~Li and G.~Cao, ``{Providing Privacy-Aware Incentives for Mobile Sensing},''
  in \emph{Proc. of the 11th IEEE PerCom}, 2013.

\bibitem{Nisan2007}
N.~Nisan, T.~Roughgarden, E.~Tardos, and V.~V. Vazirani, \emph{Algorithmic Game
  Theory}.\hskip 1em plus 0.5em minus 0.4em\relax Cambridge University Press,
  2007.

\bibitem{Cormen2001}
T.~Cormen, C.~Leiserson, R.~Rivest, and C.~Stein, \emph{Introduction to
  Algorithms}, 2nd~ed.\hskip 1em plus 0.5em minus 0.4em\relax MIT Press and
  McGraw-Hill, 2001.

\bibitem{Myerson1981}
R.~Myerson, ``{Optimal Auction Design},'' \emph{Mathematics of Operations
  Research}, vol.~6, no.~1, pp. 58--73, 1981.

\bibitem{Archer2001}
A.~Archer and E.~Tardos, ``{Truthful Mechanisms for One-Parameter Agents},'' in
  \emph{Proc. of the 42th IEEE FOCS}, 2001, pp. 482--491.

\bibitem{Dynkin1963}
E.~B. Dynkin, ``{The Optimum Choice of the Instant for Stopping a Markov
  Process},'' \emph{Soviet Math. Dokl.}, vol.~4, pp. 238--240, 1963.

\bibitem{Lenstra1990}
J.~Lenstra, D.~Shmoys, and E.~Tardos, ``{Approximation Algorithms for
  Scheduling Unrelated Parallel Machines},'' \emph{Mathematical Programming},
  vol.~46, no.~3, pp. 259--271, 1990.

\bibitem{Christodoulou2010}
G.~Christodoulou and A.~Kov\'{a}cs, ``{A Deterministic Truthful PTAS for
  Scheduling Related Machines},'' in \emph{Proc. of the 21th ACM-SIAM SODA},
  2010, pp. 1005--1016.

\bibitem{Dhangwatnotai2011}
P.~Dhangwatnotai, S.~Dobzinski, S.~Dughmi, and T.~Roughgarden, ``{Truthful
  Approximation Schemes for Single-Parameter Agents},'' \emph{SIAM Journal on
  Computing}, vol.~40, no.~3, pp. 915--933, 2011.

\bibitem{Koutsoupias2013}
E.~Koutsoupias and A.~Vidali, ``{A Lower Bound of 1 + {\ensuremath{\phi}} for
  Truthful Scheduling Mechanisms},'' \emph{Algorithmica}, vol.~66, no.~1, pp.
  211--223, 2013.

\bibitem{Karlin2005}
A.~Karlin, D.~Kempe, and T.~Tamir, ``{Beyond VCG: Frugality of Truthful
  Mechanisms},'' in \emph{Proc. of the 46th IEEE FOCS}, 2005, pp. 615--626.

\bibitem{Singer2010}
Y.~Singer, ``{Budget Feasible Mechanisms},'' in \emph{Proc. of 51th IEEE FOCS},
  2010, pp. 765--774.

\bibitem{Chen2011}
N.~Chen, N.~Gravin, and P.~Lu, ``{On the Approximability of Budget Feasible
  Mechanisms},'' in \emph{Proc. of the 22th ACM-SIAM SODA}, 2011, pp. 685--699.

\bibitem{Badanidiyuru2012}
A.~Badanidiyuru, R.~Kleinberg, and Y.~Singer, ``{Learning on a Budget: Posted
  Price Mechanisms for Online Procurement},'' in \emph{Proc. of the 13th ACM
  EC}, 2012, pp. 128--145.

\bibitem{Garey1990}
M.~R. Garey and D.~S. Johnson, \emph{Computers and Intractability; A Guide to
  the Theory of NP-Completeness}.\hskip 1em plus 0.5em minus 0.4em\relax W. H.
  Freeman \& Co., 1990.

\end{thebibliography}

\appendix
%
% We only sketch the proofs given limited space.
%
\begin{IEEEproof}[\textbf{Proof of Theorem}~\ref{thm:nphard}]
  We prove the NP-hardness of the MCS problem by a reduction from the Partition problem~\cite{Garey1990}. Given a set of $n$ integers $S=\{a_1,a_2,...,a_n\}$, the Partition problem is to decide whether the set $S$ can be partitioned into two subsets such that the sum of the numbers in one subset equals the sum of the numbers in another. Suppose that there are $n$ tasks and $n$ users in the MCS problem, and each user $A_i$ can perform only task $K_i$. Let the length of the available time period of any user be $1$, and let $a_i=d_i=\mu_i$ for any $i$. Let the budget $G=\frac{1}{2}\sum_{i=1}^n a_i$. The MCS decision problem asks if the owner can obtain a revenue $R \ge G$. Obviously, this problem is equivalent to the Partition problem on the set $S$. Since the Partition problem is NP-complete, the MCS problem is NP-hard.
\end{IEEEproof}

\begin{IEEEproof}[\textbf{Proof of Theorem}~\ref{thm:correctness}]
  As the users' strategic behaviours are not considered here, it can be easily seen by line~\ref{ln:ir} in \textbf{Algorithm~\ref{alg:assc}} that any user can always get a payment no less than his sensing cost. Hence we only need to prove that the total amount paid to the users is no more than the budget $G$. For any $1\leq i\leq h$ we have
  \begin{eqnarray}
    R(\vec{y}_{i})&=& R(\vec{y}_{{i-1}})+|y_{n_i}|\mu_{n_i} \nonumber\\
    &\leq& R(\vec{y}_{{i-1}})+\mu_{n_i}\left(\frac{G}{2d_{n_i}}-\frac{R(\vec{y}_{{i-1}})}{\mu_{n_i}}\right)\nonumber\\
    &=&{\mu_{n_i}G}/({2d_{n_i}}), \nonumber
  \end{eqnarray}
  hence we get
  \begin{eqnarray}
    {\mu_{n_i}}/{d_{n_i}} &\geq& {2R(\vec{y}_{i})}/{G}, \label{eqn:xinjia}
  \end{eqnarray}
  and
  \begin{eqnarray}
    {\mu_{n_1}}/{d_{n_1}}\geq {\mu_{n_2}}/{d_{n_2}}\geq \cdots \geq {\mu_{n_h}}/{d_{n_h}}\geq {2R(\vec{y})}/{G}.\nonumber
  \end{eqnarray}
  So the total amount paid to the users is
  \begin{eqnarray}
    \sum_{1\leq i\leq h} d_{n_i}\cdot |y_{n_i}|~~\leq~~\sum_{1\leq i\leq h} \mu_{n_i}\cdot |y_{n_i}|\cdot \frac{G}{2R(\vec{y})}~~=~~\frac{G}{2}. \label{eqn:totallesshalfg}
  \end{eqnarray}
  Therefore, \textbf{Algorithm~\ref{alg:assc}} yields a feasible solution.
\end{IEEEproof}

\begin{IEEEproof}[\textbf{Proof of Theorem}~\ref{thm:conditionalar}]
Suppose that \textbf{Algorithm}~\ref{alg:assc} has $l$ effective iterations if we replace line~\ref{ln:allocatetime} by
\begin{eqnarray}
q\leftarrow \min\left\{|Z_j|,  \frac{G-\sum_{i=1}^n d_i*|w_i(\vec{b})|}{d_j}\right\},
\end{eqnarray}
and let $\vec{x}_i$ be the current vector $\vec{w}(\vec{b})$ after the $i$th effective iteration ($1\leq i\leq l$) is executed in this case. Clearly, $h\leq l$, and the user scheduled in the $i$th effective iteration under this case can also be denoted by $A_{n_i}: 1 \leq i\leq l$. Let $R'(\vec{x}_i)=\sum_{j=1}^i \mu_{n_j}|x_{n_j}|$.

Let $\mathcal{X}_0=\emptyset$ and $\mathcal{X}_i=\{A_{n_1},A_{n_2},...,A_{n_i}\}$ for any $1\leq i\leq l$. From \textbf{Algorithm}~\ref{alg:assc} we know that, for any $1\leq j< l$ and any $1\leq i\leq j$, $T_{n_i}$ must be covered by $\vec{x}_j$, i.e., $T_{n_i}\subseteq \bigcup_{1\leq \ell\leq j\wedge \kappa_{n_\ell}=\kappa_{n_i}} f_{n_\ell}$ where $\vec{f}=\vec{x}_j$. Therefore, we have
%For any $1\leq j<l$, let $Y$ be the set of participants that are scheduled by the optimal solution but are not scheduled in $\vec{x}_j$, i.e., $Y=\mathcal{A}-\{A_{n_1},A_{n_2},...,A_{n_j}\}$.
%%$Y=\{A_i\in \mathcal{A}| |w^*_i|\neq 0 \wedge |r_i|=0 \}$.
%From \textbf{Algorithm}~\ref{alg:assc} we know that $T_{n_i}$ is ``covered'' by $\vec{x}_j$ for any $1\leq i\leq j$, i.e., $T_{n_i}\subseteq \bigcup_{1\leq f\leq j\wedge \kappa_{n_f}=\kappa_{n_i}} x_{n_f}$. Hence, when $0\leq j<l$, we have
\begin{eqnarray}
&&R(\vec{w}^*)-R'(\vec{x}_j)\leq \sum\nolimits_{i:A_i\in \mathcal{A}\backslash \mathcal{X}_j} |w^*_{i}|\mu_{i}\nonumber\\
&=& \sum\nolimits_{i:A_i\in \mathcal{A}\backslash \mathcal{X}_j} |w^*_{i}|d_i\cdot \frac{\mu_{i}}{d_{i}} \leq G\cdot \frac{\mu_{n_{j+1}}}{d_{n_{j+1}}}  \label{eqn:greedyislarge} \\
&=& G\cdot \frac{|x_{n_{j+1}}|\mu_{n_{j+1}}}{|x_{n_{j+1}}|d_{n_{j+1}}}=G\cdot\frac{R'(\vec{x}_{j+1})-R'(\vec{x}_j)}{|x_{n_{j+1}}|d_{n_{j+1}}} \label{eqn:recursioneqn1}
\end{eqnarray}
where (\ref{eqn:greedyislarge}) holds because of the greedy selection rule in line~\ref{ln:greedysel}. This yields
\begin{eqnarray}
R'(\vec{x}_{j+1})&\geq& \frac{|x_{n_{j+1}}|d_{n_{j+1}}}{G}R(\vec{w}^*)\nonumber \\
&&+\left(1-\frac{|x_{n_{j+1}}|d_{n_{j+1}}}{G}\right)R'(\vec{x}_{j}) \label{eqn:recursioneqn2}
\end{eqnarray}
Note that equation~(\ref{eqn:recursioneqn1}) and (\ref{eqn:recursioneqn2}) also hold for $j=0$ since $R'(\vec{x}_0)=0$. Therefore, when $k=1$, we have:
\begin{eqnarray}
R'(\vec{x}_k)\geq \left[1-\prod_{i=1}^k \left(1-\frac{|x_{n_i}|d_{n_i}}{G}\right)\right]R(\vec{w}^*) \label{eqn:rxkislarge}
\end{eqnarray}
By induction and using equation~(\ref{eqn:recursioneqn2}), for any $1<k\leq l$, we also have
\begin{eqnarray}
&&R'(\vec{x}_{k})
%&\geq& \frac{|x_{n_{k+1}}|b_{n_{k+1}}}{G}R(\vec{w}^*)+(1-\frac{|x_{n_{k+1}}|b_{n_{k+1}}}{G})R(\vec{x}_{k})\\
\geq \frac{|x_{n_{k}}|d_{n_{k}}}{G}R(\vec{w}^*) \nonumber\\
&&+\left(1-\frac{|x_{n_{k}}|d_{n_{k}}}{G}\right)\left[1-\prod_{i=1}^{k-1} \left(1-\frac{|x_{n_i}|d_{n_i}}{G}\right)\right]R(\vec{w}^*) \nonumber\\
&&=\left[1-\prod_{i=1}^{k} \left(1-\frac{|x_{n_i}|d_{n_i}}{G}\right)\right]R(\vec{w}^*), \nonumber
\end{eqnarray}
which means that equation~(\ref{eqn:rxkislarge}) holds for any $1\leq k\leq l$.

Now we assume that $\sum_{i=1}^l|x_{n_i}|d_{n_i}=G$. In this case, using equation~(\ref{eqn:rxkislarge}) we can get:
\begin{eqnarray}
%R(\vec{x})&=& R(\vec{x}_l)\nonumber\\
R'(\vec{x}_l)&\geq& \left[1-\prod\nolimits_{j=1}^l \left(1-\frac{{|x_{n_j}|d_{n_j}}}{G}\right)\right]R(\vec{w}^*)\nonumber\\
&=& \left[1-\prod_{j=1}^l \left(1-\frac{|x_{n_j}|d_{n_j}}{\sum_{i=1}^l|x_{n_i}|d_{n_i}}\right)\right]R(\vec{w}^*)\nonumber\\
&\geq& \left(1-(1-{1}/{l})^l\right)R(\vec{w}^*)\nonumber\\
&\geq& (1-{1}/\mathbf{e})R(\vec{w}^*) \label{eqn:xislarge}
\end{eqnarray}

On the other side, if $\sum_{i=1}^l|x_{n_i}|d_{n_i}<G$, then we must have $T_j\subseteq \bigcup_{1\leq i\leq l}x_{n_i}$ for any $j\notin \{n_1,n_2,...,n_l\}$, because otherwise $|x_j|\neq 0$, which is a contradiction. This implies that $R'(\vec{x}_l)\geq R(\vec{w}^*)$. Consequently, we know that equation~(\ref{eqn:xislarge}) always holds.

%This implies that $\vec{x}_l$ is an optimal solution to the MCS problem in this case and hence we have $R'(\vec{x}_l)=R(\vec{w}^*)$. Consequently, we know that equation~(\ref{eqn:xislarge}) always holds.

It can be seen that $|y_{n_i}|= |x_{n_i}|$ for any $1\leq i\leq h-1$ and $|y_{n_h}|\leq |x_{n_h}|$. From line~\ref{ln:sel} we know $|y_{n_h}|=\lfloor\frac{G}{2d_{n_h}}-\frac{R(\vec{y}_{{h-1}})}{\mu_{n_h}}\rfloor$. Hence we get
%It can be seen that the users assigned with non-empty schedules by \textbf{Algorithm}~\ref{alg:assc} for $\beta =0$ are $A_{n_1},A_{n_2},...A_{n_h}$.
%Suppose that there are $h$ loops in lines 3-13 of Algorithm 1 which selects participants (i.e., executes line 11) when $\beta =0$. Clearly, $h\leq l$ and the participants selected by \textbf{Algorithm}~\ref{alg:assc} for $\beta =0$ are $A_{n_1},A_{n_2},...A_{n_h}$. Since $x_{n_h}$ can be fractional, we know that
\begin{eqnarray}
|x_{n_h}|\geq \frac{G}{2d_{n_h}}-\frac{R(\vec{y}_{{h-1}})}{\mu_{n_h}}-1 = \frac{G}{2d_{n_h}}-\frac{R'(\vec{x}_{{h-1}})}{\mu_{n_h}}-1\nonumber
\end{eqnarray}
This yields
\begin{eqnarray}
R'(\vec{x}_{h})=R'(\vec{x}_{{h-1}})+|x_{n_h}|\mu_{n_h} \geq {G\cdot \mu_{n_h}}/({2d_{n_h}})-\mu_{n_h}\nonumber
\end{eqnarray}
%\begin{eqnarray}
%R(\vec{x}_{h})&=&R(\vec{x}_{{h-1}})+|x_{n_h}|\mu_{n_h} \geq {G\cdot \mu_{n_h}}/({2d_{n_h}})\nonumber
%\end{eqnarray}
Therefore, for any $h\leq i\leq l$ we have
\begin{eqnarray}
{\mu_{n_i}}/{d_{n_i}}\leq {\mu_{n_h}}/{d_{n_h}}\leq {2(R'(\vec{x}_{h})+\mu_{n_h})}/{G} \label{eqn:selrule}
\end{eqnarray}
Since $R'(\vec{x}_{{h-1}})\leq R(\vec{y})$, using equation~(\ref{eqn:selrule}) we get:
\begin{eqnarray}
R'(\vec{x}_l)&=&R'(\vec{x}_{{h-1}}) + \sum_{i=h}^l \mu_{n_i}|x_{n_i}|\nonumber\\
&\leq& R(\vec{y}) +\frac{2(R'(\vec{x}_{h})+\mu_{n_h})}{G}\cdot \sum_{i=h}^l d_{n_i}|x_{n_i}|\nonumber\\
&\leq& R(\vec{y}) + 2(R'(\vec{x}_{h-1})+\mu_{n_h}|x_{n_h}|)+2\mu_{n_h}\nonumber\\
&\leq& 3R(\vec{y})+4 \mu_{n_h}|x_{n_h}|\nonumber\\
&\leq& 3R(\vec{y})+(1-1/{\mathbf{e}}-4\epsilon)R(\vec{w}^*)\nonumber
\end{eqnarray}
Combing this with equation~(\ref{eqn:xislarge}), the theorem follows.
\end{IEEEproof}

\begin{IEEEproof} [\textbf{Proof of Lemma}~\ref{lma:largevalueisless}]
  Because $\mu_i/d_i'\leq \mu_i/d_i$, bidding $b'_i$ can only postpone the schedule assignment for $A_i$ according to line~\ref{ln:greedysel} of \textbf{Algorithm~\ref{alg:assc}}. Hence the length of the time period scheduled for $A_i$ can only decrease when $A_i$ bids $b_i'$, due to line~\ref{ln:allocatetime} of \textbf{Algorithm~\ref{alg:assc}}.
\end{IEEEproof}

\begin{IEEEproof}[\textbf{Proof of Lemma}~\ref{lma:smallintervalisless}]
Let $\widehat{{Z}}$ and $\widehat{{Z}}'$ be the uncovered available time period of $A_i$ when $A_i$ is scheduled by \textbf{Algorithm~\ref{alg:assc}} with the input bids being $(b_i,b_{-i})$ and $(b_i',b_{-i})$, respectively. As $[s'_i,e'_i]\subseteq [s_i,e_i]$, we must have $\widehat{{Z}}'\subseteq \widehat{{Z}}$. Hence the lemma follows due to lines~\ref{ln:allocatetime} and \ref{ln:sel} of \textbf{Algorithm~\ref{alg:assc}}.
  %The proof is similar to the proof of \textit{\textbf{Lemma}~\ref{lma:largevalueisless}}, hence is omitted.
\end{IEEEproof}

\begin{IEEEproof} [\textbf{Proof of Lemma}~\ref{lma:biddingbound}]
  Suppose that $j=n_c\!:1\leq c\leq h$ (hence $j\in \{n_1,n_2,...,n_h\}$) and $d_{j}> \frac{\mu_{j}\cdot G}{R(\vec{y})}$ by contradiction, we have
  \begin{eqnarray}
    %&\leq& \sum_{c\leq i\leq h}|y_{n_i}|\mu_{n_i} + \sum_{1\leq i\leq c-1}|y_{n_i}|\mu_{n_i} -R(\vec{y}_{c-1})\nonumber\\
    \!\!\!\!\!\!\!R(\vec{y})-R(\vec{y}_c) &\leq& \sum\nolimits_{c\leq i\leq h}|y_{n_i}|\cdot d_{n_i}\cdot ({\mu_{n_i}}/{d_{n_i}}) \nonumber\\
    &\leq& \frac{\mu_j}{d_j}\cdot \sum_{1\leq i\leq h}|y_{n_i}|\cdot d_{n_i}~~\leq~~\frac{G}{2}\cdot \frac{\mu_j}{d_j} \label{eqn:lesshalfg}\\
    &<& {R(\vec{y})}/{2},\nonumber
  \end{eqnarray}
  where (\ref{eqn:lesshalfg}) holds due to (\ref{eqn:totallesshalfg}). Therefore $R(\vec{y})< 2R(\vec{y}_c)$. On the other side, (\ref{eqn:xinjia}) suggests ${\mu_{j}}/{d_{j}}\geq {2R(\vec{y}_c)}/{G}$. Combining these inequalities yields $d_{j}<\frac{\mu_{j}\cdot G}{R(\vec{y})}$; a contradiction. % which is a contradiction to our assumption. Hence the lemma follows.%\label{eqn:contra2}%> \frac{R(\vec{y})}{G}.
\end{IEEEproof}

\begin{IEEEproof}[\textbf{Proof of Theorem}~\ref{thm:correctmch}]
  It is easy to see that lines~\ref{ln:bstart}-\ref{ln:bend} of \textbf{Algorithm~\ref{alg:mch}} can output a feasible solution satisfying IR to the MCS problem. The output of lines~\ref{ln:callassc}-\ref{ln:endcal} satisfies IR according to \textit{\textbf{Theorem}~\ref{thm:payment}}. Hence we only need to prove that $\sum_{j\in \mathcal{W}} p_{j}(\vec{b})\leq G$ for $\mathcal{W}=\{n_1,n_2,...,n_h\}$. According to \textit{\textbf{Lemma}~\ref{lma:biddingbound}}, no user $A_j\!: j\in \mathcal{W}$ can bid $(d_j,s_j,e_j)$ with $d_{j}>\frac{\mu_{j}\cdot G}{R(\vec{y})}$, because otherwise he/she will get an empty schedule. Therefore, using \textit{\textbf{Theorem}~\ref{thm:payment}} and \textit{\textbf{Lemma}~\ref{lma:largevalueisless}} we can get
  \begin{eqnarray}
    && p_j(b_j,b_{-j}) \nonumber\\
    &=& d_j\cdot |y_j(b_j,b_{-j})| + \int_{d_j}^{\infty} |y_j\left((v,s_j,e_j),b_{-j}\right)|\mathrm{d}v \nonumber\\
    &=& d_j\cdot |y_j(b_j,b_{-j})| + \int_{d_j}^{\frac{\mu_{j}\cdot G}{R(\vec{y})}} |y_j\left((v,s_j,e_j),b_{-j}\right)|\mathrm{d}v \nonumber\\
    &\leq& d_j\cdot |y_j(b_j, b_{-j})| + \left(\frac{\mu_{j}\cdot G}{R(\vec{y})}-d_j\right)\cdot |y_j(b_j,b_{-j})|\nonumber\\
    &=& \left({\mu_{j}\cdot G}/{R(\vec{y})}\right)\cdot |y_j(b_j,b_{-j})|. \nonumber
  \end{eqnarray}
  Given that $R(\vec{y})=\sum_{j\in \mathcal{W}}\mu_j\cdot |y_j(b_j,b_{-j})|$, we can prove $\sum_{j\in \mathcal{W}} p_{j}(\vec{b})\leq G$ by summing up $p_j(b_j,b_{-j})$ for all $j\in \mathcal{W}$, hence the theorem follows.
\end{IEEEproof}

\begin{IEEEproof}[\textbf{Proof of Theorem}~\ref{thm:mchar}]
  For any $\epsilon\in (0,\frac{\mathbf{e}-1}{4\mathbf{e}})$, if (\ref{eqn:biggest}) is satisfied, then the mechanism in \textbf{Algorithm~\ref{alg:mch}} has a revenue of at least $\frac{4}{3}\epsilon\cdot R(\vec{y}^*)$ with probability of $\frac{1}{2}$; if (\ref{eqn:biggest}) is not satisfied, then we have:
  \begin{eqnarray}
    \mu_j~~\geq~~\max_{i:d_i\leq G} (\mu_i\cdot |T_i|)/\lambda~~\geq~~\left(\frac{\mathbf{e}-1}{4\lambda \mathbf{e}}-\frac{\epsilon}{\lambda}\right)R(\vec{y}^*), \nonumber
  \end{eqnarray}
  hence the mechanism has a revenue of at least $\left(\frac{\mathbf{e}-1}{4\lambda \mathbf{e}}-\frac{\epsilon}{\lambda}\right)\cdot R(\vec{y}^*)$ with probability of $\frac{1}{2}$. Therefore, the overall approximation ratio of the mechanism is $\mathcal{O}(1)$. For example, if we set $\epsilon=\frac{3}{28}(1-1/\mathbf{e})$, then the expected revenue of the mechanism is at least $\frac{1}{7\lambda}(1-1/\mathbf{e})\cdot R(\vec{y}^*)$.
\end{IEEEproof}

\begin{IEEEproof}[\textbf{Proof of Theorem}~\ref{thm:timcom}]
  Line~\ref{ln:callassc} of \textbf{Algorithm~\ref{alg:mch}} calls \textbf{Algorithm~\ref{alg:assc}} that has a time complexity of $\mathcal{O}(n^2)$ due to the sorting of the users. Line~\ref{ln:callpay} is iterated at most $n$ times and each calculates the payment to one user by calling \textbf{Algorithm}~\ref{alg:payment} that has a time complexity of $\mathcal{O}(n)$. The time complexity of lines~\ref{ln:bstart}-\ref{ln:bend} in \textbf{Algorithm~\ref{alg:mch}} is $\mathcal{O}(n)$.  Consequently, the overall time complexity of \textbf{Algorithm~\ref{alg:mch}} is $\mathcal{O}(n^2)$.
\end{IEEEproof}

\begin{IEEEproof} [\textbf{Proof of Theorem}~\ref{thm:onlmchpartb}]
%Let $B_j$ be the event that the $j$th come agent is the best and let $C_j$ be the event that the $j$th come agent is selected. Let $D_j$ be the event that the best among the first $j(j\geq \lfloor \frac{n}{e} \rfloor)$  come agent is among the first $\lfloor \frac{n}{e} \rfloor$ come agent. Then we know that the probability of the best agent is selected is:
%\begin{eqnarray}
%&&\sum_{j=1}^n \mathrm{Prob}\{C_j|B_j\}\cdot \mathrm{Prob}\{B_j\}\nonumber\\
%&=&\sum_{j=\lfloor \frac{n}{e} \rfloor+1}^n \mathrm{Prob}\{C_j|B_j\}\cdot \mathrm{Prob}\{B_j\}\nonumber\\
%&=&\sum_{j=\lfloor \frac{n}{e} \rfloor+1}^n \mathrm{Prob}\{D_{j-1}|B_j\}\cdot \mathrm{Prob}\{B_j\}\nonumber\\
%&\geq& \sum_{j=\lfloor \frac{n}{e} \rfloor+1}^n \frac{\lfloor \frac{n}{e} \rfloor}{j-1}\cdot \frac{1}{n}\nonumber\\
%&=& ({\lfloor \frac{n}{e} \rfloor}/n)\cdot\sum_{j=\lfloor \frac{n}{e} \rfloor}^{n-1} \frac{1}{j}\geq ({\lfloor \frac{n}{e} \rfloor}/n)\int_{\lfloor \frac{n}{e} \rfloor}^{n} \frac{1}{v}\mathrm{d}v \nonumber\\
%&=&({\lfloor \frac{n}{e} \rfloor}/n)\ln ({n}/{\lfloor \frac{n}{e} \rfloor})\nonumber\\
%&\geq& -(1/e-1/n)\ln (1/e-1/n)\nonumber\\
%&\geq& 1/e-1/n \geq 1/e-1/3
%\end{eqnarray}
%where holds because $1/e\geq {\lfloor \frac{n}{e} \rfloor}/n\geq 1/e-1/n$ and the function $v\ln (1/v)$ is increasing when $v\in (0,1/e]$.
%
  Similar to the secretary algorithm~\cite{Dynkin1963}, we can prove that the user $A_k$ is selected with probability of at least $1/\mathbf{e}-1/3$, where $k=\arg\max_{i:d_i\leq G}\mu_i$, as far as there are more than two users (i.e., $n \ge 3$). Hence \textbf{Algorithm~\ref{alg:secra}} has a constant competitive ratio of $1/(150\lambda)$ with probability of at least $1/\mathbf{e}-1/3$.
\end{IEEEproof}

\begin{IEEEproof} [\textbf{Proof of Lemma}~\ref{lma:deltabound}]
  Let $\{Y_1,Y_2,...,Y_n\}$ be a set of independent random variables such that  $Y_{i}=|y^*_{i}|\cdot \mu_{i}$ if $i\in \{\sigma_1,\sigma_2,...,\sigma_{\xi}\}$, and $Y_{i}=0$ otherwise. Clearly, $Y_i\leq \Lambda$ for any $1\leq i\leq n$. Let $Y=\sum_{i=1}^n Y_{i}$. Hence, $\Delta_1=Y$ and $\Delta_2=R(\vec{y}^*)-Y$. For any $1\leq i\leq n$, we have
  \begin{eqnarray}
    &&\mathrm{Prob}\left(i\in\{\sigma_1,\sigma_2,\cdots,\sigma_{\xi}\}\right) \nonumber\\
    &=& \sum_{j=1}^n \mathrm{Prob}(\sigma_j=i)\cdot \mathrm{Prob}(\xi\geq j)~=~\frac{1}{n}\mathbb{E}(\xi)~=~\frac{1}{2}. \nonumber
  \end{eqnarray}
  So we know $\mathbb{E}(Y) = \sum_{i=1}^n \mathbb{E}(Y_i) = \frac{1}{2}\sum_{i=1}^n |y^*_{i}|\cdot \mu_{i} = \frac{R(\vec{y}^*)}{2}$. According to the Chernoff bound, we get
  \begin{eqnarray}
    \mathrm{Prob}\left(\Delta_1\leq \frac{R(\vec{y}^*)}{3}\right)=\mathrm{Prob}\left(Y\leq \left(1-\frac{1}{3}\right)\mathbb{E}(Y) \right) \nonumber\\
    \leq \mathbf{e}^{\frac{-\frac{1}{9}\cdot \mathrm{E}(Y)}{2\Lambda}} = \mathbf{e}^{-\frac{R(\vec{y}^*)}{36\Lambda}} \leq \mathbf{e}^{-\frac{150}{36}}\leq 0.016, \nonumber
  \end{eqnarray}
and
\begin{eqnarray}
&&\mathrm{Prob}\left(\Delta_2\leq \frac{R(\vec{y}^*)}{4}\right)=\mathrm{Prob}\left(Y\geq \left(1+\frac{1}{2}\right)\mathbb{E}(Y) \right) \nonumber\\
%&=& \mathrm{Prob}\left(\Delta_1\geq (1+\frac{1}{2})\mathbb{E}(\Delta_1) \right)\nonumber\\
&\leq& \left(\frac{\mathbf{e}^{\frac{1}{2}}}{(1+\frac{1}{2})^{(1+\frac{1}{2})}} \right)^\frac{\mathrm{E}(Y)}{\Lambda}\leq (0.9)^\frac{R(\vec{y}^*)}{2\Lambda} \leq 0.9^{75}\leq 0.001. \nonumber
\end{eqnarray}
By the union bound, we know
%\begin{eqnarray}
%\mathrm{Prob}\{\Delta_1\leq {R(\vec{w}^*)}/{3}\vee \Delta_2\leq {R(\vec{w}^*)}/{4}\}\leq 0.017 \nonumber
%\end{eqnarray}
%Hence
\begin{eqnarray}
\mathrm{Prob}\{\Delta_1\geq {R(\vec{y}^*)}/{3}\wedge \Delta_2\geq {R(\vec{y}^*)}/{4}\}\geq 0.983. \nonumber
\end{eqnarray}
So the lemma follows.
\end{IEEEproof}

\begin{IEEEproof} [\textbf{Proof of Lemma}~\ref{lma:boundapx}]
Let $opt_1$ be the revenue of the optimal solution for the first arrived $\xi$ users $\{A_{\sigma_1},...,A_{\sigma_{\xi}}\}$. Using \textit{\textbf{Theorem}~\ref{thm:conditionalar}} with $\epsilon=\frac{\mathbf{e}-1}{4\mathbf{e}}-\frac{1}{150}$, we get
\begin{eqnarray}
R(\vec{r})~~\geq~~\frac{4}{3}\cdot \left(\frac{\mathbf{e}-1}{4\mathbf{e}}-\frac{1}{150}\right)\cdot opt_1~~\geq~~\frac{1}{5}\cdot opt_1. \nonumber
\end{eqnarray}
As $opt_1\geq \Delta_1$, we get $R(\vec{r})\geq \Delta_1/5$. On the other hand, $R(\vec{r})\leq  opt_1\leq  R(\vec{y}^*)$, hence the lemma follows.
%As $\Delta_2\geq \frac{R(\vec{w}^*)}{4}$, we know $R(\vec{r})\leq 4\Delta_2$.
%Let $(\sigma_1,\sigma_2,...,\sigma_n)$ be a permutation of $(1,2,...,n)$ such that $|\rho_{\sigma_1}|\cdot \mu_{\sigma_1}\geq |\rho_{\sigma_2}|\cdot \mu_{\sigma_2}\geq...\geq |\rho_{\sigma_n}|\cdot \mu_{\sigma_n}$. Let $\{Y_1,Y_2,...,Y_n\}$ be a set of random variables such that $Y_{\sigma_i}=|\rho_{\sigma_i}|\cdot \mu_{\sigma_i}$ if $1\leq \sigma_i\leq \xi$, and $Y_{\sigma_i}=0$ otherwise. Let $Y=\sum_{i=1}^k Y_{i}$ and $\bar{Y}=R(\vec{w}^*)-Y$. So we have $Y=\sum_{i=1}^\xi |\rho_{i}|\cdot \mu_i$ and $\bar{Y}=\sum_{i=\xi+1}^n |\rho_{i}|\cdot \mu_i$
\end{IEEEproof}

\begin{IEEEproof} [\textbf{Proof of Lemma}~\ref{lma:onlmchparta}]
  Let $O_1=\{\sigma_i|\xi+1\leq i\leq n \bigwedge y^*_{\sigma_i}\not\subseteq \bigcup_{j:\kappa_{j}=\kappa_{\sigma_i}}y_{j}\}$ and $O_2=\{\sigma_i|\xi+1\leq i\leq n \bigwedge y^*_{\sigma_i}\subseteq \bigcup_{j:\kappa_{j}=\kappa_{\sigma_i}}y_{j}\}$. Since $\sum\nolimits_{i\in O_2} |y^*_{i}|\cdot \mu_{i} \leq R(\vec{y})$, we have
  %
%Let $O_1$ be the set of users' ids that are scheduled in $\Delta_2$ but not in \textbf{Algorithm}~\ref{alg:onlinemch}, i.e., $O_1=\{\sigma_i|\xi+1\leq i\leq n \wedge |w^*_{\sigma_i}|\neq 0 \wedge |z_{\sigma_i}|= 0\}$. Let $O_2$ be the set of users' ids that are both scheduled in \textbf{Algorithm}~\ref{alg:onlinemch} and $\Delta_2$, i.e., $O_2=\{\sigma_i|\xi+1\leq i\leq n \wedge |w^*_{\sigma_i}|\neq 0 \wedge |z_{\sigma_i}|\neq 0\}$. Let $O_3$ be the set of of users' ids scheduled in $\Delta_2$ but are covered by the schedule of \textbf{Algorithm}~\ref{alg:onlinemch}, i.e., $O_3=\{\sigma_i|\xi+1\leq i\leq n \wedge w^*_{\sigma_i}\subseteq \bigcup_{j:\kappa_{\sigma_i}=\kappa_{j}}z_{j}\}$
%Then we know that for any $i\in O_2$ we have  $w^*_{i}\subseteq \bigcup_{j:\kappa_{i}=\kappa_{j}}z_{j}$, which means
%\begin{eqnarray}
%\sum\nolimits_{i\in O_2} |w^*_{i}|\cdot \mu_{i} \leq \sum\nolimits_{i=\xi+1}^{n} |z_{\sigma_i}|\cdot \mu_{\sigma_i} \nonumber
%\end{eqnarray}
  %
  \begin{eqnarray}
    \Delta_2-R(\vec{y}) &=& \sum_{i\in O_1} |y^*_{i}|\cdot \mu_{i}+\sum_{i\in O_2} |y^*_{i}|\cdot \mu_{i}-R(\vec{y})\nonumber\\
    &\leq& \sum\nolimits_{i\in O_1} |y^*_{i}|\cdot \mu_{i}.\nonumber
  \end{eqnarray}
  %
%\begin{eqnarray}
%&&\Delta_2-\sum\nolimits_{i=\xi+1}^n |z_{\sigma_i}|\cdot \mu_{\sigma_i}\nonumber\\
%&=&\sum_{i\in O_1} |w^*_{i}|\cdot \mu_{i}+\sum_{i\in O_2} |w^*_{i}|\cdot \mu_{i}-\sum_{i=\xi+1}^n |z_{\sigma_i}|\cdot \mu_{\sigma_i}\nonumber\\
%&\leq& \sum\nolimits_{i\in O_1} |w^*_{i}|\cdot \mu_{i}\nonumber
%\end{eqnarray}
%$O=\{\sigma_i|\xi+1\leq i\leq n \wedge |w^*_{\sigma_i}|\neq 0 \wedge |z_{\sigma_i}|= 0\wedge w^*_{\sigma_i}\not\subseteq \bigcup_{\sigma_j:\kappa_{\sigma_i}=\kappa_{\sigma_j}}z_{\sigma_j}\}$.
  %
  For any $i\in O_1$, we must have $|y^*_i|\neq 0$, $|y_i|=0$, and $|F_i|\geq1$. %Note that we have $|F_i|\geq1$ for any $i\in O_1$.
  %
%in line~\ref{ln:allocationjudge}.
%otherwise the whole $T_i$ is covered by $\vec{z}$.
  %
  So we can discuss line~\ref{ln:allocationjudge} as follows:

  Case 1: Suppose that $d_i>\eta=5G\cdot\mu_i /R(\vec{r})$  for any $i\in O_1$. Using \textit{\textbf{Lemma}~\ref{lma:boundapx}} we have:
  \begin{eqnarray}
    \Delta_2-R(\vec{y}) &\leq& \sum\nolimits_{i\in O_1} |y^*_{i}|\cdot d_i \cdot {R(\vec{r})}/{(5G)}\nonumber\\
    &\leq& {R(\vec{r})}/{5}~~\leq~~R(\vec{y}^*)/5.\nonumber
    \end{eqnarray}
%\begin{eqnarray}
%&&\Delta_2-\sum_{i=\xi+1}^n |z_{\sigma_i}|\cdot \mu_{\sigma_i}\leq \sum_{i\in O_1} |w^*_{i}|\cdot d_i \cdot \frac{R(\vec{r})}{5G}\nonumber\\
%&\leq& {R(\vec{r})}/{5} \leq R(\vec{w}^*)/5\nonumber
%\end{eqnarray}
  %
  As $R(\vec{y}^*)\leq 4\Delta_2$ according to \textit{\textbf{Lemma}~\ref{lma:deltabound}}, we have
  \begin{eqnarray}
    R(\vec{y})~~\geq~~\Delta_2 -R(\vec{y}^*)/5~~\geq~~R(\vec{y}^*)/20.\nonumber
  \end{eqnarray}
%\begin{eqnarray}
%\sum_{i=\xi+1}^n |z_{\sigma_i}|\cdot \mu_{\sigma_i}\geq \Delta_2 -\frac{1}{5}R(\vec{w}^*)\geq \frac{1}{20}R(\vec{w}^*)\nonumber
%\end{eqnarray}

  Case 2: Suppose that there exists $i\in O_1$ such that $d_i\leq 5G\cdot\mu_i /R(\vec{r})$ but $\eta|F_i|=5|F_i|\cdot G\cdot\mu_i /R(\vec{r})>M$. In this case, using \textit{\textbf{Lemma}~\ref{lma:deltabound}} and \textit{\textbf{Lemma}~\ref{lma:boundapx}} we get:
  \begin{eqnarray}
    M &\leq& 5|F_i|\cdot G\cdot\mu_i /R(\vec{r})~~\leq~~5G\cdot \Lambda /R(\vec{r})\nonumber\\
    &\leq& 5G\cdot \frac{R(\vec{y}^*)}{150}\bigg{/} \left(\frac{\Delta_1}{5}\right)~~\leq~~\frac{G}{2},
  \end{eqnarray}
  hence
  \begin{eqnarray}
    \frac{G}{2}&\leq& \sum_{i=\xi+1}^n p_{\sigma_i}~~\leq~~\sum_{i=\xi+1}^n |y_{\sigma_i}|\cdot 5G\cdot\mu_{\sigma_i} /R(\vec{r}) \nonumber\\
    &=& 5G\cdot R(\vec{y})/R(\vec{r})~~\leq~~25G\cdot R(\vec{y})/\Delta_1 \nonumber\\
    &\leq& 75G\cdot R(\vec{y})/R(\vec{y}^*),
  \end{eqnarray}
  which yields $R(\vec{y})\geq \frac{1}{150}R(\vec{y}^*)$. Therefore, lines~\ref{ln:ranstart}-\ref{ln:ranend} of \textbf{Algorithm~\ref{alg:onlinemch}} has a competitive ratio of $1/150$ with probability of at least 0.983.
\end{IEEEproof}

\begin{proof} [\textbf{Proof of Theorem}~\ref{thm:artimcom}]
  Similar to the proof of \textit{\textbf{Theorem}~\ref{thm:mchar}}, it can be easily proven that \textbf{Algorithm~\ref{alg:onlinemch}} has an $\mathcal{O}(1)$ competitive ratio based on \textit{\textbf{Lemma}~\ref{lma:onlmchparta}} and \textit{\textbf{Theorem}~\ref{thm:onlmchpartb}}. \textbf{Algorithm}~\ref{alg:onlinemch} has its running time predominantly spent on line~\ref{ln:calransample}, which has a $\mathcal{O}(n^2)$ worst-case time complexity.
\end{proof}

%\section*{Acknowledgment}
%The authors would like to thank...\cite{Fudenberg1991}\cite{Yang2012,Mun2009,Thiagarajan2009,krishna2002,Consolvo2008,Miluzzo2008,Khan2013}\cite{Archer2001,Kearns2002,Lau2011,Osborne1994,Garey1990,Yuan2009}
%\bibliography{MyLib}
%\bibliographystyle{alpha}
%\bibliographystyle{apacite}

\end{document}